\documentstyle[aps,prd,eqsecnum,amssymb,amsbsy,epsfig,floats]{revtex}

\newcommand{\be}{\begin{equation}}
\newcommand{\ee}{\end{equation}}
\newcommand{\bea}{\begin{eqnarray}}
\newcommand{\eea}{\end{eqnarray}}
\newcommand{\bs}{\boldsymbol}
\newcommand{\dst}{\displaystyle}
\newcommand{\pa}{\partial}

\newcommand{\dsub}{d_{\text{sub}}}
\newcommand{\F}{{\mathcal F}}
\newcommand{\Fo}{{\mathcal F}_o}
\newcommand{\Fsub}{{\mathcal F}_{\text{sub}}}
\newcommand{\tmin}{\tau_{\text{min}}}

\def\tav#1{\left\langle#1\right\rangle}

\begin{document}

\title{Data analysis of gravitational-wave signals
from spinning neutron stars.\\ IV.\ An all-sky search}

\author{Pia Astone}
\address{Istituto Nazionale di Fisica Nucleare,
(INFN)-Rome I, 00185 Rome, Italy}

\author{Kazimierz M.\ Borkowski}
\address{Center for Astronomy,
Nicolaus Copernicus University,
Gagarina 11, 87-100 Toru\'n, Poland}

\author{Piotr Jaranowski}
\address{Institute of Theoretical Physics,
University of Bia{\l}ystok,
Lipowa 41, 15-424 Bia{\l}ystok, Poland}

\author{Andrzej Kr\'olak}
\address{Institute of Mathematics,
Polish Academy of Sciences,
\'Sniadeckich 8, 00-950 Warsaw, Poland}

\maketitle

\begin{abstract}
We develop a set of data analysis tools for a realistic all-sky search for
continuous gravitational-wave signals.  The methods that we present apply to
data from both the resonant bar detectors that are currently in operation and
the laser interferometric detectors that are in the final stages of construction
and commissioning.  We show that with our techniques we shall be able to perform
an all-sky 2-day long coherent search of the narrow-band data from the resonant
bar EXPLORER with no loss of signals with the dimensionless amplitude greater
than $2.8\times10^{-23}$.

\vspace{2ex}\noindent
PACS number(s): 95.55.Ym, 04.80.Nn, 95.75.Pq, 97.60.Gb
\end{abstract}

\section{Introduction and summary}

It was shown that an all-sky full frequency bandwidth coherent search of data of
many months duration for continuous gravitational-wave signals is
computationally too prohibitive to be realized with presently available
computing power \cite{BCCS98,JKS98}.  Consequently two alternative approaches
emerged.  One that we shall call {\em tracking} involves tracking of lines in
the time-frequency plane built from the Fourier transforms of short (around 40
minutes long) stretches of data \cite{PAFS97,PS99}.  The other that we shall
call {\em stacking} involves dividing the data into shorter (around a day long)
stretches, searching each stretch for signals, and enhancing the detectability
by incoherently summing the Fourier transforms of data stretches \cite{BC99}.
In this paper we develop data analysis tools for the first stage of the stacking
approach, namely for the coherent search of a few days long stretches of data.
These tools involve storage of the data in the Fourier domain database,
calculation of the precise position of the detector with respect to the solar
system barycenter using JPL ephemeris, calculation of the detection thresholds,
approximation of the signal by a simple linear model, and construction of the
grid of templates in the parameter space ensuring no loss of signals.

In this theoretical work we have in mind a particular set of data, namely the
data collected by the EXPLORER bar detector.  This detector is most sensitive
over certain two narrow bandwidths of about 1~Hz wide at frequency around 1~kHz.
To make the search computationally feasible we propose an all-sky search of a
few days long stretches of data in the narrow band where the detector has the
best sensitivity.  We find that with a $250$ Mflops of computing power we can
carry out a complete all-sky search in around a month for signals within a
bandwidth of 0.76~Hz at frequency of 922~Hz and for observation time of 2 days.
We assume that the minimum characteristic time for the change of the signal's
frequency is 1000 years.  By our search we shall be able to detect all the
continuous gravitational-wave signals with the parameters given above and with
the dimensionless amplitude larger than $2.8\times10^{-23}$ with 99\%
confidence.

The tools developed in this work can be applied not only to the analysis of the
bar data but also to the interferometer data.  In the case of the wide-band
interferometer one can divide the data into many narrow bands and analyze each
band using the algorithms presented in this paper.

The plan of the paper is as follows.  In Sec.\ II we present responses
of both a laser interferometer and a resonant bar to a continuous gravitational
wave.  The optimal data processing method is derived in Sec.\ \ref{Sec:OPT}.  In
Sec.\ \ref{Sec:FDDB} we describe the construction of the {\em frequency domain
database} and in Sec.\ \ref{T2B} we discuss how to calculate the precise
position of the detector with respect to the solar system barycenter.  In Sec.\
\ref{Sec:LM} we introduce an approximate model of the detector's response called
the {\em linear} model and we describe its performance.  In Sec.\ \ref{Sec:G} we
construct a grid of templates in the parameter space ensuring no loss of
signals, we calculate the computational requirements to do the search, and we
perform Monte Carlo simulations to test the performance of our grid and our
search algorithm.  In Sec.\ \ref{Sec:EXPs} we discuss and choose parameters for
our planned all-sky search of the EXPLORER detector data.  Several details are
presented in the Appendixes.

Many methods presented in this paper rely on general analytic tools developed in
a previous paper of this series \cite{JK00} that we shall call Paper III.

\section{Detector's noise-free response}

In this section we present the response of a laser interferometer and a resonant
bar detector to a continuous gravitational wave.  We include the effects of both
amplitude and frequency modulation of the response.

Dimensionless noise-free response function $h$ of the gravitational-wave
detector to a weak plane gravitational wave in the long wavelength approximation
[i.e.\ when the size of the detector is much smaller than the reduced wavelength
$\lambda/(2\pi)$ of the wave] can be written as a linear combination of the {\em
wave polarization functions} $h_+$ and $h_\times$:
\be
\label{h}
h(t) = F_+(t)\,h_+(t) + F_{\times}(t)\,h_{\times}(t),
\ee
where $F_+$ and $F_\times$ are called the {\em beam-pattern functions}.

\subsection{Interferometric beam-pattern functions}

Noise-free response function $h$ of the laser interferometric detector is 
defined as the difference between the wave induced relative length changes of 
the two interferometer arms. Derivation of formula (\ref{h}) in the case of the 
laser interferometer can be found e.g.\ in Sec.~II~A of Ref.\ \cite{JK94}.

{\em Explicit} formulas for the interferometric beam-pattern functions $F_+$ and 
$F_\times$ from Eq.\ (\ref{h}) are derived e.g.\ in Sec.~II~A of Ref.\ 
\cite{JKS98}. The functions read
\begin{mathletters}
\label{bpi}
\bea
F_+(t) &=& \sin\zeta\left[a(t)\cos2\psi+b(t)\sin2\psi\right],
\\[2ex]
F_\times(t) &=& \sin\zeta\left[b(t)\cos2\psi-a(t)\sin2\psi\right],
\eea
\end{mathletters}
where
\begin{mathletters}
\label{abi}
\bea
\label{ai}
a(t) &=&
\frac{1}{4} \sin2\gamma (1 + \sin^2\phi) (1 + \sin^2\delta)
\cos[2(\alpha-\phi_r-\Omega_r t)]
\nonumber\\[2ex]&&
- \frac{1}{2} \cos2\gamma \sin\phi (1 + \sin^2\delta)
\sin[2(\alpha-\phi_r-\Omega_r t)]
\nonumber\\[2ex]&&
+ \frac{1}{4} \sin2\gamma \sin2\phi \sin2\delta
\cos(\alpha-\phi_r-\Omega_r t)
\nonumber\\[2ex]&&
- \frac{1}{2} \cos2\gamma \cos\phi \sin2\delta
\sin(\alpha-\phi_r-\Omega_r t)
\nonumber\\[2ex]&&
+ \frac{3}{4} \sin2\gamma \cos^2\phi \cos^2\delta,
\\[2ex]
\label{bi}
b(t) &=&
\cos2\gamma \sin\phi \sin\delta \cos[2(\alpha-\phi_r-\Omega_r t)]
\nonumber\\[2ex]&&
+ \frac{1}{2} \sin2\gamma (1 + \sin^2\phi) \sin\delta
\sin[2(\alpha-\phi_r-\Omega_r t)]
\nonumber\\[2ex]&&
+ \cos2\gamma \cos\phi \cos\delta \cos(\alpha-\phi_r-\Omega_r t)
\nonumber\\[2ex]&&
+ \frac{1}{2} \sin2\gamma \sin2\phi \cos\delta
\sin(\alpha-\phi_r-\Omega_r t).
\eea
\end{mathletters}
In Eqs.\ (\ref{bpi}) $\zeta$ is the angle between the interferometer arms 
(usually $\zeta=90^\circ$) and $\psi$ is the polarization angle of the wave.  In
Eqs.\ (\ref{abi}) the angles $\alpha$ and $\delta$ are respectively right
ascension and declination of the gravitational-wave source.  The geodetic
latitude of the detector's site is denoted by $\phi$ [its precise definition is
given in Eq.\ (\ref{geodphi}) below], the angle $\gamma$ determines the
orientation of the detector's arms with respect to local geographical
directions:  $\gamma$ is measured counter-clockwise from East to the bisector of
the interferometer arms.  The rotational angular velocity of the Earth is
denoted by $\Omega_r$, and $\phi_r$ is a deterministic phase which defines the
position of the Earth in its diurnal motion at $t=0$ (the sum $\phi_r+\Omega_r
t$ essentially coincides with the local sidereal time of the detector's site,
i.e.\ with the angle between the local meridian and the vernal point; see Sec.\
VII~B below).

\subsection{Bar beam-pattern functions}

Derivation of {\em explicit} formulas for the bar beam-pattern functions $F_+$ 
and $F_\times$ from Eq.\ (\ref{h}) is relegated to Appendix~A of the present 
paper. The functions read
\begin{mathletters}
\label{bpb}
\bea
F_+(t) &=& a(t)\cos2\psi+b(t)\sin2\psi,
\\[2ex]
F_\times(t) &=& b(t)\cos2\psi-a(t)\sin2\psi,
\eea
\end{mathletters}
where
\begin{mathletters}
\label{abb}
\bea
a(t) &=&
\frac{1}{2} (\cos^2\gamma-\sin^2\gamma\sin^2\phi) (1+\sin^2\delta)
\cos[2(\alpha-\phi_r-\Omega_r t)]
\nonumber\\[2ex]&&
+ \frac{1}{2} \sin2\gamma \sin\phi (1+\sin^2\delta)
\sin[2(\alpha-\phi_r-\Omega_r t)]
\nonumber\\[2ex]&&
- \frac{1}{2} \sin^2\gamma \sin2\phi \sin2\delta
\cos[\alpha-\phi_r-\Omega_r t]
\nonumber\\[2ex]&&
+ \frac{1}{2} \sin2\gamma \cos\phi \sin2\delta \sin(\alpha-\phi_r-\Omega_r t)
\nonumber\\[2ex]&&
+ \frac{1}{2} (1-3 \sin^2\gamma \cos^2\phi) \cos^2\delta,
\\[2ex]
b(t) &=&
-\sin2\gamma \sin\phi \sin\delta \cos[2(\alpha-\phi_r-\Omega_r t)]
\nonumber\\[2ex]&&
+ (\cos^2\gamma-\sin^2\gamma\sin^2\phi) \sin\delta
\sin[2(\alpha-\phi_r-\Omega_r t)]
\nonumber\\[2ex]&&
- \sin2\gamma \cos\phi \cos\delta \cos(\alpha-\phi_r-\Omega_r t)
\nonumber\\[2ex]&&
- \sin^2\gamma \sin2\phi \cos\delta \sin(\alpha-\phi_r-\Omega_r t).
\eea
\end{mathletters}
In Eqs.\ (\ref{bpb}) and (\ref{abb}) $\alpha$, $\delta$, $\psi$, $\phi$, 
$\Omega_r$, and $\phi_r$ all have the same meaning as in Eqs.\ 
(\ref{bpi})--(\ref{abi}); the angle $\gamma$ determines the orientation of the 
bar detector with respect to local geographical directions: $\gamma$ is 
measured counter-clockwise from East to the bar's axis of symmetry.

\subsection{Wave polarization functions}

We are interested in a continuous gravitational wave, which is described by the 
wave polarization functions of the form
\begin{mathletters}
\label{wavepol}
\bea
h_+(t) &=& h_{0+} \cos\Psi(t),
\\[2ex]
h_\times(t) &=& h_{0\times} \sin\Psi(t),
\eea
\end{mathletters}
where $h_{0+}$ and $h_{0\times}$ are the independent amplitudes of the two wave 
polarizations. These amplitudes depend on the physical mechanisms generating
the gravitational wave. In the case of a wave originating from a spinning 
neutron star these amplitudes can be estimated by
\be
\label{hon}
h_o = 4.23\times10^{-23}\left(\frac{\epsilon}{10^{-5}}\right)
\left(\frac{I}{10^{45}~\mbox{g cm}^2}\right)
\left(\frac{\mbox{1~kpc}}{r}\right)\left(\frac{f}{\rm 1~kHz}\right)^2,
\ee 
where $I$ is the neutron star moment of inertia with respect to the rotation
axis, $\epsilon$ is the poloidal ellipticity of the star, and $r$ is the
distance to the star.  The value of $10^{-5}$ of the parameter $\epsilon$ in the
above estimate should be treated as an upper bound.  In reality it may be
several orders of magnitude less.

The phase $\Psi$ of the wave is given by
\begin{mathletters}
\label{phaza}
\bea
\label{phaza1}
\Psi(t) &=& \Phi_0 + \Phi(t),
\\[2ex]
\label{phaza2}
\Phi(t) &=& \sum_{k=0}^{s_1} \omega_k \frac{t^{k+1}}{(k+1)!}
+ \frac{{\bf n}_0\cdot{\bf r}_{\rm SSB}(t)}{c}
\sum_{k=0}^{s_2} \omega_k \frac{t^k}{k!}.
\eea
\end{mathletters}
In Eqs.\ (\ref{phaza}) $\Phi_0$ is the initial phase of the wave form,
$\bf{r}_{\rm SSB}$ is the vector joining the solar system barycenter (SSB) with
the detector, ${\bf n}_0$ is the constant unit vector in the direction from the
SSB to the neutron star.  We assume that the gravitational wave form is almost
monochromatic around some angular frequency $\omega_0$ which we define as
instantaneous angular frequency evaluated at the SSB at $t=0$, $\omega_k$
($k=1,2,\ldots$) is the $k$th time derivative of the instantaneous angular
frequency at the SSB evaluated at $t=0$.  To obtain formulas (\ref{phaza}) we
model the frequency of the signal in the rest frame of the neutron star by a
Taylor series.  For the detailed derivation of the phase model (\ref{phaza}) see
Sec.\ II~B and Appendix A of Ref.\ \cite{JKS98}.  The integers $s_1$ and $s_2$
are the number of spin downs to be included in the two components of the phase.
We need to include enough spin downs so that we have a sufficiently accurate
model of the signal to extract it from the noise.  This will depend on a given
data and in particular on the length of the observation time.

\subsection{A linear representation}

It is convenient to write the response of gravitational-wave detectors described 
above in the following form [cf.\ Eqs.\ (\ref{h}), (\ref{bpi}), (\ref{bpb}), 
(\ref{wavepol}), and (\ref{phaza1})]
\be
\label{sig}
h(t) = \sum^{4}_{i=1} A_{i}\,h_{i}(t), 
\ee
where the four constant amplitudes $A_{i}$ are given by
\begin{mathletters}
\bea
A_1 &=& \sin\zeta \left( h_{0+}\cos2\psi\cos\Phi_0
- h_{0\times}\sin2\psi\sin\Phi_0 \right),
\\[2ex]
A_2 &=& \sin\zeta \left( h_{0\times}\cos2\psi\sin\Phi_0
+ h_{0+}\sin2\psi\cos\Phi_0 \right),
\\[2ex]
A_3 &=& \sin\zeta \left( -h_{0+}\cos2\psi\sin\Phi_0
- h_{0\times}\sin2\psi\cos\Phi_0 \right),
\\[2ex]
A_4 &=& \sin\zeta \left( h_{0+}\cos2\psi\cos\Phi_0
- h_{0\times}\sin2\psi\sin\Phi_0 \right).
\eea
\end{mathletters}
The amplitudes $A_i$ depend on the physical mechanisms generating the 
gravitational wave.

The time dependent functions $h_i$ have the form
\begin{mathletters}
\bea
h_1(t) &=& a(t) \cos\Phi(t),
\\[2ex]
h_2(t) &=& b(t) \cos\Phi(t),
\\[2ex]
h_3(t) &=& a(t) \sin\Phi(t),
\\[2ex]
h_4(t) &=& b(t) \sin\Phi(t),
\eea
\end{mathletters}
where the functions $a$ and $b$ are given by Eqs.\ (\ref{abi}) for an
interferometer and by Eqs.\ (\ref{abb}) for a bar, and $\Phi$ is the phase given
by Eq.\ (\ref{phaza2}).  The modulation amplitudes $a$ and $b$ depend on the
right ascension $\alpha$ and the declination $\delta$ of the source (they also
depend on the angles $\phi$ and $\gamma$).  The phase $\Phi$ depends on the
frequency $\omega_0$, $s$ spin-down parameters $\omega_k$ $(k=1,\ldots,s)$, and
on the angles $\alpha$, $\delta$.  We call parameters $\omega_0$, $\omega_k$,
$\alpha$, $\delta$ the {\em phase parameters}.  Moreover the phase $\Phi$
depends on the latitude $\phi$ of the detector.  The whole signal $h$ depends on
$s+5$ unknown parameters:  $h_{0+}$, $h_{0\times}$, $\alpha$, $\delta$,
$\omega_0$, $\omega_k$.  The formulas above apply to both resonant bar and
laser interferometric detectors, but for the case of bars the angle $\zeta$ is
always $90^\circ$.  The main feature of the above representation is that the
signal is decomposed into a linear combination of several components.

\section{Optimal data analysis method}
\label{Sec:OPT}

The signal given by Eq.\ (\ref{sig}) will be buried in the noise of a detector.
Thus we are faced with the problem of detecting the signal and estimating its
parameters.  A standard method is the method of {\em maximum likelihood (ML)
detection} that consists of maximizing the likelihood function, which we shall
denote by $\Lambda$, with respect to the parameters of the signal.  If the
maximum of $\Lambda$ exceeds a certain threshold calculated from the false alarm
probability that we can afford, we say that the signal is detected.  The values
of the parameters that maximize $\Lambda$ are said to be the {\em maximum
likelihood estimators} of the parameters of the signal.  The magnitude of the
maximum of $\Lambda$ determines the probability of detection of the signal.

We assume that the noise $n$ in the detector is an additive, stationary, 
Gaussian, and zero-mean continuous random process. Then the data $x$ (if the 
signal $h$ is present) can be written as
\be
x(t) = n(t) + h(t).
\ee
The log likelihood function has the form
\be
\label{loglr}
\log\Lambda = (x|h) - \frac{1}{2}(h|h),
\ee
where the scalar product $(\,\cdot\,|\,\cdot\,)$ is defined by 
\be
\label{SP}
(x|y) := 4\, \Re \int^{\infty}_{0}
\frac{\tilde{x}(f)\tilde{y}^{*}(f)}{S_h(f)} df.
\ee
In Eq.\ (\ref{SP}) tilde denotes the Fourier transform, asterisk means complex 
conjugation, and $S_h$ is the {\em one-sided} spectral density of the detector's 
noise.

Since we assume that over the bandwidth of the signal $h$ the spectral density 
$S_h(f)$ is nearly constant and equal to $S_h(f_0)$, where $f_0$ is the 
frequency of the signal measured at the SSB at $t=0$, the scalar products in 
Eq.\ (\ref{loglr}) can be approximated by 
\begin{mathletters}
\label{approx1}
\bea
(x|h) &\approx& \frac{2}{S_h(f_0)} \int^{T_o}_{0} x(t)\,h(t)\,dt,
\\[2ex]
(h|h) &\approx& \frac{2}{S_h(f_0)} \int^{T_o}_{0} \left[h(t)\right]^2\,dt,
\eea
\end{mathletters}
where $T_o$ is the observation time, and the observation interval is 
$\left[0,T_o\right]$. It is useful to introduce the following notation 
\be
\tav{x} := \frac{1}{T_o} \int^{T_o}_{0} x(t)\,dt.
\ee
Applying this notation and making use of Eqs.\ (\ref{approx1}), the log 
likelihood ratio from Eq.\ (\ref{loglr}) can be written as 
\be
\label{loglr3}
\ln\Lambda \approx \frac{2T_o}{S_h(f_0)}
\left( \tav{xh}-\frac{1}{2}\tav{h^2} \right).
\ee

In Sec.\ III of Paper III we have analyzed in detail the likelihood ratio for
the general case of a signal consisting of $N$ narrow-band components.  Here we
only summarize the results of Paper III and adapt them to the case of our signal
(\ref{sig}).  The signal $h$ depends linearly on four amplitudes $A_i$.  The
likelihood equations for the ML estimators $\widehat{A}_{i}$ of the amplitudes
$A_i$ are given by
\be
\label{ampest}
\frac{\partial\ln\Lambda}{\partial A_{i}} = 0,\quad i=1,\ldots,4.
\ee
One can easily find the explicit analytic solution to Eqs.\ (\ref{ampest}). To 
simplify formulas we assume that {\em the observation time $T_o$ is an integer 
multiple of one sidereal day}, i.e., $T_o=n(2\pi/\Omega_r)$ for some positive 
integer $n$. Then the time average of the product of the functions $a$ and $b$ 
vanishes, $\tav{ab}=0$, and the analytic formulas for the ML estimators of the 
amplitudes are given by
\begin{mathletters}
\label{amle0}
\bea
\widehat{A}_1 &\approx& 2 \frac{\tav{x h_1}}{\tav{a^2}},
\\[2ex]
\widehat{A}_2 &\approx& 2 \frac{\tav{x h_2}}{\tav{b^2}},
\\[2ex]
\widehat{A}_3 &\approx& 2 \frac{\tav{x h_3}}{\tav{a^2}},
\\[2ex]
\widehat{A}_4 &\approx& 2 \frac{\tav{x h_4}}{\tav{b^2}}.
\eea
\end{mathletters}
Explicit formulas for the time averages $\tav{a^2}$ and $\tav{b^2}$ are given in 
Appendix B. The reduced log likelihood function $\F$ is the log likelihood 
function where amplitude parameters $A_i$ were replaced by their estimators 
$\widehat{A}_i$. By virtue of Eqs.\ (\ref{amle0}) from Eq.\ (\ref{loglr3}) one 
gets 
\be
\label{OS}
\F \approx \frac{2}{S_h(f_0) T_o}
\left( \frac{|F_a|^2}{\tav{a^2}} + \frac{|F_b|^2}{\tav{b^2}} \right),
\ee
where
\begin{mathletters}
\label{Fab}
\bea
F_{a} &:=& \int^{T_o}_0 x(t)\, a(t) \exp[-i\Phi(t)]\,dt,
\\[2ex]
F_{b} &:=& \int^{T_o}_0 x(t)\, b(t) \exp[-i\Phi(t)]\,dt.
\eea
\end{mathletters}
We thus see that the ML detection in Gaussian noise leads to a {\em matched
filter}.  The ML estimators of the signal's parameters are obtained in two
steps.  Firstly, the estimators of the frequency, the spin-down parameters, and
the angles $\alpha$ and $\delta$ are obtained by maximizing the functional
$\F$ with respect to these parameters.  Secondly, the estimators of
the amplitudes $A_{i}$ are calculated from the analytic formulas (\ref{amle0})
with the correlations $\tav{xh_i}$ evaluated for the values of the parameters
obtained in the first step.  Thus filtering for the narrow-band
gravitational-wave signal requires {\em two linear complex}
filters.  The amplitudes $A_i$ of the signal depend on the physical mechanisms
generating gravitational waves.  If we know these mechanisms and consequently we
know the dependence of $A_i$ on a number of parameters we can estimate these
parameters from the estimators of the amplitudes by least-squares method.

We shall now summarize the statistical properties of the functional $\F$.  We 
first assume that the phase parameters are known and consequently
that $\F$ is only a function of random data $x$ (the case when the phase
parameters are unknown and need to be estimated will be considered later in this
section).  When the signal is absent $2\F$ has a $\chi^2$ distribution
with four degrees of freedom and when the signal is present it has a noncentral
$\chi^2$ distribution with four degrees of freedom and noncentrality parameter
equal to the \emph{optimal signal-to-noise ratio} $d$ defined as
\be
\label{snr}
d := \sqrt{(h|h)}.
\ee
This is the maximum signal-to-noise ratio that can be achieved for a signal in
additive noise with a {\em linear} filter \cite{Da}.  This fact does not depend
on the statistics of the noise. Consequently the probability density functions 
$p_0$ and $p_1$ when respectively the signal is absent and present are given by 
\begin{mathletters}
\bea
\label{p0}
p_0(\F) &=& \F \exp(-\F),
\\[2ex]
\label{p1}
p_1(d,\F) &=& \frac{\sqrt{2\F}}{d} I_1\Big(d\sqrt{2\F}\Big)
\exp\left(-\F-\frac{1}{2}d^2\right),
\eea
\end{mathletters}
where $I_1$ is the modified Bessel function of the first kind and order 1.
The false alarm probability $P_F$ is the probability that $\F$ exceeds a certain 
threshold $\Fo$ when there is no signal. In our case we have 
\be
\label{PF}
P_F(\Fo) := \int_{\Fo}^\infty p_0(\F)\, d\F = (1 + \Fo) \exp(-\Fo).
\ee
The probability of detection $P_D$ is the probability that $\F$ exceeds the
threshold $\Fo$ when the signal-to-noise ratio is equal to $d$: 
\be
\label{PD}
P_D(d,\Fo) := \int_{\Fo}^{\infty} p_1(d,\F)\, d\F.
\ee
The integral in the above formula cannot be evaluated in terms of known
special functions. We see that when the noise in the detector is Gaussian
and the phase parameters are known the probability of detection of the
signal depends on a single quantity: the optimal signal-to-noise ratio $d$.

When the phase parameters are unknown the optimal statistics $\F$
depends not only on the random data $x$ but also on the phase parameters that we
shall denote by $\bs{\xi}$.  Such an object is called in the theory of
stochastic processes a {\em random field}.  Let us consider the correlation
function of the random field
\be
\label{covdef}
C(\bs{\xi},\bs{\xi}') := \text{E} \left\{ [\F(\bs{\xi})-m(\bs{\xi})]
[\F(\bs{\xi}')-m(\bs{\xi}')] \right\},
\ee
where
\be
m(\bs{\xi}) := \text{E} \left\{\F(\bs{\xi})\right\}.
\ee
When the correlation function $C$ depends only on the difference 
$\bs{\xi}-\bs{\xi}'$ the random field $\F$ is called {\em homogeneous} and for 
such fields we have developed in Paper III a theory to calculate the false alarm 
probabilities.

The main idea is to divide the space of the phase parameters $\bs{\xi}$ into
{\em elementary cells} which boundary is determined by the \emph{characteristic
correlation hypersurface} of the random field $\F$.  The correlation
hypersurface is defined by the requirement that at this hypersurface the
correlation $C$ equals half of the maximum value of $C$.  Assuming that $C$
attains its maximum value when $\bs{\xi}-\bs{\xi}'=0$ the equation of the
characteristic correlation hypersurface reads
\be
\label{gcov1}
C({\bs\tau}) = \frac{1}{2} C(0),
\ee
where we have introduced $\bs{\tau}:=\bs{\xi}-\bs{\xi}'$. Let us expand 
the left-hand side of Eq.\ (\ref{gcov1}) around $\bs{\tau}=0$ up to terms of 
second order in $\bs{\tau}$. We arrive at the equation 
\be
\label{gcov2}
\sum_{i,j=1}^M G_{ij} \tau_i\tau_j = 1,
\ee
where $M$ is the dimension of the parameter space and the matrix $G$ is
defined as follows 
\be
\label{gmatrix}
G_{ij} := -\frac{1}{C(0)} 
\frac{\pa^2{C(\bs{\tau})}}{\pa{\tau_i}\partial{\tau_j}} 
\Bigg\vert_{\bs{\tau}=0}.
\ee
Equation (\ref{gcov2}) defines an $M$-dimensional hyperellipsoid which we take
as an approximation to the characteristic correlation hypersurface of our
random field and we call the \emph{correlation hyperellipsoid}. This
approximation is helpful in establishing upper limit estimates of the number
of elementary cells in the parameter space. The $M$-dimensional Euclidean
volume $V_{\text{cell}}$ of the hyperellipsoid defined by Eq.\ (\ref{gcov2})
equals 
\be
\label{vc}
V_{\text{cell}} = \frac{\pi^{M/2}}{\Gamma(M/2+1)\sqrt{\det G}},
\ee
where $\Gamma$ denotes the Gamma function.

We estimate the number $N_c$ of elementary cells by dividing the total
Euclidean volume $V_{\text{total}}$ of the parameter space by the volume 
$V_{\text{cell}}$ of the correlation hyperellipsoid, i.e.\ we have 
\be
\label{NT}
N_c = \frac{V_{\text{total}}}{V_{\text{cell}}}.
\ee

We approximate the probability distribution of $\F(\bs{\xi})$ in each cell by
the probability distribution $p_0(\F)$ when the parameters are known [in our
case it is given by Eq.\ (\ref{p0})].  The values of the statistics $\F$ in each
cell can be considered as independent random variables.  The probability that
$\F$ does not exceed the threshold $\Fo$ in a given cell is $1-P_F(\Fo)$, where
$P_F(\Fo)$ is given by Eq.\ (\ref{PF}).  Consequently the probability that $\F$
does not exceed the threshold $\Fo$ in \emph{all} the $N_c$ cells is
$[1-P_F(\Fo)]^{N_c}$.  The probability $P^T_F$ that $\F$ exceeds $\Fo$ in
\emph{one or more} cell is thus given by
\be
\label{FP}
P^T_F(\Fo) = 1 - [1 - P_F(\Fo)]^{N_c}.
\ee
This is the false alarm probability when the phase parameters are unknown.
The expected number of false alarms $N_F$ is given by 
\be
\label{NF}
N_F = N_c\, P_F(\Fo).
\ee
By means of Eqs.\ (\ref{PF}) and (\ref{NT}), Eq.\ (\ref{NF}) can be rewritten as 
\be
\label{NFi}
N_F = \frac{V_{\text{total}}}{V_{\text{cell}}} (1 + \Fo) \exp(-\Fo).
\ee

Using Eq.\ (\ref{NF}) we can express the false alarm probability $P^T_F$
from Eq.\ (\ref{FP}) in terms of the expected number of false alarms. Using
$\lim_{n\rightarrow\infty}(1+\frac{x}{n})^n=\exp(x)$ we have that for large
number $N_c\gg1$ of cells 
\be
\label{FPap}
P^T_F({\mathcal F}_o) \approx 1 - \exp(-N_F).
\ee
When the expected number $N_F$ of false alarms is small, $N_F\ll1$, we have 
$P^T_F(\Fo)\approx N_F$.

It will very often be the case that the filter we use to extract the signal from
the noise is not optimal.  This may be the case when we do not know the exact
form of the signal (this is almost always the case in practice) or we choose a
suboptimal filter to reduce the computational cost and simplify the analysis.
For such a case in Paper III we have developed a theory of suboptimal filtering.
The suboptimal statistics $\Fsub$ has the same form as the optimal statistics
$\F$ and the data analysis procedure consists of maximizing the value of the
suboptimal statistics.  We have that
\be
\label{EF}
\text{E}_1\{\Fsub\} = 1 + \frac{1}{2}\, \dsub^2, 
\ee
where $\text{E}_1$ is the expectation value when the signal is present and 
the parameter $\dsub$ is the suboptimal signal-to-noise ratio.

The formulas for the false alarm and detection probabilities have the same form
as in the case of the optimal filter except that the optimal signal-to-noise 
ratio 
is replaced by the suboptimal one.  Also for the suboptimal filter the number of
cells $N_{sc}$ may be different than for the optimal one.  Thus the false alarm
probability is given by
\be
\label{sFP}
P^T_{sF}(\Fo) = 1 - [1 - P_F(\Fo)]^{N_{sc}}.
\ee
The expected number of false alarms $N_{sF}$ reads 
\be
\label{sNF}
N_{sF} = N_{sc}\, P_F(\Fo).
\ee
The detection probability takes the form
\be
\label{sDP}
P_D(\dsub,\Fo) := \int_{\Fo}^{\infty} p_1(\dsub,\F)\, d\F,
\ee
where the probability density function $p_1$ is given by Eq.\ (\ref{p1}). 
Equation (\ref{EF}) implies that maximizing expectation value of the suboptimal 
statistics when the signal is present is equivalent to maximizing suboptimal 
signal-to-noise ratio. We can identify the maximum $d_{s\text{max}}$ of $\dsub$ 
as the signal-to-noise ratio of the suboptimal filter. It was shown in Paper III 
that
\be
d_{s\text{max}} = \text{FF}\, d,
\ee
where FF is the {\em fitting factor} introduced by Apostolatos \cite{A1}. The 
fitting factor FF between a signal $h(t;\bs{\theta})$ and a filter 
$h'(t;\bs{\theta}')$ ($\bs{\theta}$ and $\bs{\theta}'$ are the parameters of the 
signal and the filter, respectively) is defined as 
\be
\label{ff1}
\text{FF} := \max_{\bs{\theta}'}
\frac{\biglb( h(t;{\bs\theta}) \vert h'(t;{\bs\theta}') \bigrb)}
{ \sqrt{ \biglb( h(t;{\bs\theta}) \vert h(t;{\bs\theta}) \bigrb) }
\sqrt{ \biglb( h'(t;{\bs\theta}') \vert h'(t;{\bs\theta}') \bigrb) } }.
\ee
The fitting factor is a good measure of the quality of the suboptimal filter
however the performance of the filter can only be fully determined by
the false alarm and detection probabilities given by Eqs.\ (\ref{sFP}) and 
(\ref{sDP}).

\section{The frequency domain database}
\label{Sec:FDDB}

The data collected by the detector constitute a time series with a sampling
interval $\Delta t$.  This time domain sequence can also be stored in the
frequency domain without any loss of information.  Organization of the data in
the frequency domain in a suitable way provides a very flexible database that is
very useful both for search of the data for gravitational-wave signals and for
characterization of the noise of the detector.  The first step to construct {\em
frequency domain database} consists simply of taking FFT of $2N$ data points and
storing the first $N$ points of the FFT.  To improve the quality of the
time-domain data that will eventually be extracted from the database the
original data are windowed and overlapped.  Each FFT is then normalized and
calibrated where the calibration process takes into account the transfer
function of the detector so that the units of the FFT are
strain$/\sqrt{\text{Hz}}$.  Moreover the quality of the data represented by FFT
is characterized so that it is possible to choose thresholds for vetoing the
data or criteria to weight them.  Both the calibration and the data
characterization information are stored in the header that is attached to each
FFT.  It is very important to choose a suitable number $N$ of data in each FFT.
This depends on the type of signal search that one wants to perform.  In the
case of the search for continuous sources which is a subject of this work we
choose $N$ in such a way the maximum expected Doppler shift due to the motion of
the detector with respect to SSB is less than the width of one bin.  In the case
of the EXPLORER data this leads to $N=2^{16}$.  For sampling time
$\Delta{t}=0.18176$~s this means length of data for each FFT of around 2/3 of an
hour.  Another important criterion for the choice of the length of data interval
for each FFT is that within the interval the noise is stationary.  In the case
of the EXPLORER detector stationarity is in general preserved over 2/3 of an
hour intervals.

The detailed steps to create the frequency domain database are as follows.
\begin{itemize}
\item
Each FFT is computed using $2N$ data, sampled with sampling time $\Delta{t}$. 
The data are windowed, in the time domain, before the Fourier transform. We use 
a Hamming window, that is the data $y_i$ are multiplied by 
$w_i=A-B\cos{i}+C\cos{2i}$, where $i=(0,2N-1)\cdot2\pi/(2N-1)$, $A=0.54$, 
$B=0.46$, and $C=0$. 
\item 
The FFTs are calibrated, normalized, and stored in units of 
strain/$\sqrt{\text{Hz}}$ so that their squared modulus is the spectrum. 
\item 
The basic FFTs of the database overlap for half their length.  The time 
duration of each FFT is $t_0=2N\Delta{t}$, and a new FFT is done after time 
$t_0/2$. This is important since it avoids distortions in the final time domain 
sequence --- this is the well-known ``overlap-add'' method, described in many 
data analysis textbooks. 
\end{itemize}

Once we have a database we can extract from it a time domain sequence of time 
duration $2MN\Delta{t}$ and bandwidth $\Delta\nu$ by the following procedure.
\begin{itemize}
\item
Take the data from $n'=N\Delta\nu/B$ bins in the frequency band $\Delta\nu$ of 
the actual search, where $B$ is the total bandwidth of the detector.
\item
Build a complex vector that has the following structure: 
\begin{itemize}
\item
the first datum equal to zero;
\item
the next $n'$ data equal to those from the selected bins of the FFT;
\item
zeroes from bins $n'+1$ up to the nearest subsequent bin numbered with a power 
of two; let us say that in this way one has obtained $n$ bins;
\item
zeroes in the next $n$ bins; one thus ends up with a vector that is $2n$ long.
\end{itemize}
\item
After the bandwidth has been selected, the data (still in the frequency domain)
should be windowed, to avoid edge effects in the transformed data.
\item 
Take the inverse FFT of the vector obtained above. This is a complex time series
that is called the {\em analytic signal} because by construction above its 
spectrum is zero for negative frequencies. The signal has the bandwidth
$\Delta\nu$ that is shifted towards the lower frequencies and the signal 
is sampled at a sampling rate lower by a factor $N/n$ compared 
to the original time data.\footnote{
The construction of the analytic signal is a standard procedure of lowpass 
filtering for a bandpass process. By the fact that the analytic signal is zero 
on the left frequency plan one avoids aliasing effects in the lowpass sampling 
operation; see, e.g., Ref.~\cite{tretter}.}
The time of the first sample here is exactly the same as the time
of the first datum used for the database and the total duration 
is also that of the original time stretch. 
There are fewer data because the sampling time of the original data 
is shorter. 
\item
Remove the window $w$ used when constructing the FFT database simply by 
dividing the new time domain data by $w$. This operation recovers the original 
(subsampled) time data because the only regions where the division might not 
work are the edges of the data stream where the value of the function $w$ may be 
zero, depending on the used window (a problem which, of course, has been 
overcome by overlapping of the FFTs).
\item
Repeat the steps outlined above for all the $M$ FFTs.
\item
If any FFT under consideration is vetoed or if it is missing the corresponding 
data are set to zero.
\item
Each new group of time domain data is appended to the previous groups after 
elimination of the overlapped data. Since the overlapping involved half the data 
we eliminate 1/4 of the data at the beginning and end of each stream. The first 
1/4 data in the first FFT and the last 1/4 in the last FFT can be discarded.
The data of missing or vetoed periods, set to zero as explained above, are 
appended to the other stretches in the same way.
\end{itemize}

Thus by the above procedure we have a subsampled time domain data stream which 
represents the analytic signal associated with the original data.

\section{Position and velocity of the detector with respect to the solar system
barycenter}
\label{T2B}

\subsection{Dealing with time scales}

The data acquisition system of the detector is synchronized to the Coordinated
Universal Time (UTC) as disseminated by international time services.  Thus it is
assumed that each datum corresponds to a given UTC.  The UTC scale is
essentially uniform, except for occasional 1~s steps introduced internationally
to compensate for the variable Earth rotation.  When these steps are taken into
account, the UTC is reduced to the International Atomic Time (TAI) scale, which
is uniform and differs from the Terrestrial Dynamical Time (TDT or TT, normally
used to describe celestial phenomena by astronomers) only by a constant term.
In principle, the time argument of positions of celestial objects is the
Barycentric Dynamical Time (TDB), however it differs from the TDT only by a very
small additive term (less than 0.001~s in magnitude), thus for all practical
purposes they do not have to be distinguished.  So we have:
$$
\mbox{TDB} \approx \mbox{TDT} = \mbox{TAI} + 32.184\,\mbox{s},
$$
where TAI is obtained from UTC by removing the time steps.\footnote{The time 
steps are available in the form of a table; see 
\verb|http://hpiers.obspm.fr/webiers/general/earthor/utc/|
\allowbreak\verb|table1.html|.}
One can thus relate given UTC with barycentric positions of all the major 
celestial bodies of the solar system.

However, to relate the position of a point on the Earth to the barycenter
of the Earth (the latter being obtained from the solar system ephemeris) one
has to use yet another time scale --- the rotational time scale UT1,
which is nonuniform and is determined from astronomical observations.
The difference UT1 $-$ UTC, which is maintained within $\pm0.90$ s, is taken 
from the IERS tabulations of daily values.\footnote{\label{epoc}They are 
available as \verb|eopc04.yy| files, where \verb|yy| stands for a two digit year 
number (e.g.\ \verb|99| for 1999, and \verb|00| for 2000); see
\verb|http://hpiers.obspm.fr/iers/eop/eopc04|.}

\subsection{Topocentric coordinates}

To be able to relate a point on the Earth surface to the solar system barycenter
it is necessary to know orientation of the Earth in space.  The primary effects
that should be taken into account are:  diurnal (variable) rotation, precession
and nutation of the Earth rotational axis, and polar motion.  The precession and
nutation can be accounted for by applying standard astronomical theories.  The
remaining two effects are unpredictable for a longer future, so observational
data must be used.\footnote{For past years, since 1962, the data necessary for 
reduction are included in \verb|eopc04.yy| files mentioned in footnote 
\ref{epoc}.}

The polar motion is taken into account by modifying the conventional
geographical coordinates of a point on the Earth:
\begin{mathletters}
\bea
\label{geodphi}
\phi &=& \phi_\circ + P_x \cos\lambda_\circ - P_y \sin\lambda_\circ,
\\[2ex]
\label{geodlam}
\lambda &=& \lambda_\circ + (P_x\sin\lambda_\circ
+ P_y\cos\lambda_\circ)\tan\phi_\circ,
\eea
\end{mathletters}
where $\phi_\circ$ is the conventional geographical latitude, 
$\lambda_\circ$ --- the conventional longitude, and $P_x$ and $P_y$ are the
coordinates of the pole with respect to the Conventional International Origin.

The rotational angle of the Earth is included through conversion of UTC to UT1
(the quantity UT1 $-$ UTC is tabulated in the IERS files).  UT1 serves to
calculate the apparent sidereal time $\theta$, which is essentially equal to
$\phi_r+{\Omega_r}t$ of Eqs.\ (\ref{abi}) and (\ref{abb}) plus the nutation in 
longitude projected on the celestial equator.  The sidereal time in turn is used 
to find rectangular coordinates of the point (the detector) in the IERS 
celestial reference frame:
\begin{mathletters}
\bea
\label{detc}
x_E &=& r\cos\theta,
\\[2ex]
y_E &=& r\sin\theta,
\\[2ex]
z_E &=& b\sin\psi + h\sin\phi,
\eea
\end{mathletters}
where
\be
\label{eqco}
r = a\cos\psi + h\cos\phi
\ee
is the equatorial component of the radius vector, $\psi=\arctan(b\tan\phi/a)$ is
the reduced latitude, $h$ is the height above the Earth ellipsoid, and
$a=6378.140$ km and $b=a(1-1/f)$ (where $f=298.257$) are the semiaxes of the
Earth ellipsoid.  These coordinates are affected by the polar motion through
dependence of $r$ and $\psi$ on $\phi$, and $\theta$ on $\lambda$.

The polar motion ($P_x$ and $P_y$) and UT1 $-$ UTC quantities are linearly
interpolated between the daily IERS values.

Since these coordinates are naturally referred to the epoch of date, 
they are further precessed back to the standard epoch J2000.
The precessed Cartesian coordinates $(x_E,y_E,z_E)_{2000}$ may now be
straightforwardly added to coordinates of the Earth barycenter with
respect to the solar system barycenter (which are described in the next
paragraph) to obtain the desired position vector of the detector.

\subsection{Barycentric coordinates of the Earth (JPL ephemeris)}

For computing the coordinates of the Earth barycenter, relative to the SSB, use
is made of the latest JPL Planetary and Lunar Ephemerides, ``DE405/LE405'' or
just ``DE405'',\footnote{It is available via the Internet (anonymous ftp:
\verb|navigator.jpl.nasa.gov|, the directory: \verb|ephem/export|) or on CD
(from the publisher:  Willmann-Bell, Inc.;
\verb|http://www.willbell.com/software/jpl.htm|).} created in 1997 and
described in detail by Standish \cite{S98}.  It represents an improvement over
its predecessor, DE403.  DE405 is based upon the International Celestial
Reference Frame (ICRF; an earlier popular ephemeris DE200, which has been the
basis of the {\em Astronomical Almanac} since 1984, is within 0.01 arcseconds of
the frame of the ICRF).  It constitutes of a set of Chebyshev polynomials fit
with full precision to a numerical integration over 1600 AD to 2200 AD.  Over
this interval the interpolating accuracy is not worse than 25 meters for any
planet and not worse than 1 meter for the Moon.  The JPL package allows a
professional user to obtain the rectangular coordinates of the Sun, Moon, and
nine major planets anywhere between JED 2305424.50 (1599 DEC 09) to JED
2525008.50 (2201 FEB 20).

In the application described in this paper we have used only a 21-year (1990 to
2010) subset of the original ephemeris.  The ephemeris gives separately the
position of the Earth--Moon barycenter and the Moon's position relative to this
barycenter.  The Earth position vector is thus calculated as a fraction
(involving the masses of the two bodies) of the Moon's one and opposite to the
latter.

Finally, the vector traveled by the Sun towards its apex (with the speed of 20
km/s) between J2000 and the epoch of observation is added to the Earth
barycentric position.  The direction of solar apex is assumed at 18$^{\mbox{h}}$
in right ascension and 30$^\circ$ in declination at the epoch J1900.  This
direction is precessed to J2000.  What is commonly known as the solar apex
refers rather to solar system barycenter apex.  However since the apex motion is
known only approximately, really there is no need to distinguish between motions
of the Sun and of the barycenter.

\subsection{Velocities}

Although the primary concern in this project is to convert positions, the
velocity of the gravitational-wave detector relative to the SSB may prove to be
useful in future analyzes of spectra obtained from the acquired data.

The velocity of the detector is the sum of diurnal rotational motion of the
Earth, Earth motion in space relative to the SSB, and motion of the solar system
itself towards the apex.  The last of the named components, towards the apex,
has been described in the previous paragraph.  The Earth barycentric velocity
vector is obtained directly from the JPL ephemeris along with the position
vector.  Finally, the detector motion relative to the Earth barycenter is
represented by a vector of constant length $v_\circ:=2\pi r$/(sidereal day)
directed always towards the east in the topocentric reference frame.  Thus this
diurnal velocity vector has the following Cartesian components (in the
barycentric reference frame):
\begin{mathletters}
\bea
V_x &=& v_\circ \cos(\theta+\pi/2),
\\[2ex]
V_y &=& v_\circ \sin(\theta+\pi/2),
\\[2ex]
V_z &=& 0.
\eea
\end{mathletters}
This vector is rotated back to J2000 by the precessional angle.

A short description of the FORTRAN code that generates both the position and the 
velocity of a detector with respect to the SSB is given in Appendix \ref{EPH}.

\section{A linear filter}
\label{Sec:LM}

The phase of the gravitational-wave signal contains terms arising from the
motion of the detector with respect to the SSB.  These terms consist of two
contributions, one which comes from the motion of the Earth barycenter with
respect to the SSB, and the other which is due to the diurnal motion of the
detector with respect to the Earth barycenter.  The first contribution has a
period of one year and thus for shorter observation times can be well
approximated by a few terms of the Taylor expansion.  The second term has a
period of 1 sidereal day and to a very good accuracy can be approximated by a
circular motion.  We thus propose the following {\em approximate} simple model
of the phase of the gravitational-wave signal:
\be
\label{Eq:Phs}
\Psi_s(t) = p + p_0\, t + \sum^s_{k=1} p_k\, t^{k+1} +  A \cos(\Omega_r t)
+ B \sin(\Omega_r t),
\ee
where $\Omega_r$ is the rotational angular velocity of the Earth. The parameters
$A$ and $B$ can be related to the right ascension $\alpha$ and the declination 
$\delta$ of the gravitational-wave source through the equations [cf.\ Eq.\ (18) 
in Ref.\ \cite{JKS98}]
\begin{mathletters}
\label{AB}
\bea
A &=& \frac{\omega_0 r}{c} \cos\delta \cos(\alpha - \phi_r),
\\[2ex]
B &=& \frac{\omega_0 r}{c} \cos\delta \sin(\alpha - \phi_r),
\eea
\end{mathletters}
where $\omega_0$ is the angular frequency of the gravitational-wave signal and 
$r$ is defined in Eq.\ (\ref{eqco}).

The phase model (\ref{Eq:Phs}) has the property that it is a {\em linear}
function of the parameters.  We have shown in Paper III that for linear phase
models the optimal statistics is a homogeneous random field and consequently the
statistical theory of signal detection described in Sec.\ III of the present
paper applies to this case.  The polynomial in time part of the phase $\Psi_s$
[cf.\ Eq.\ (\ref{Eq:Phs})] contains two contributions.  The first one comes from
the intrinsic frequency drift of the gravitational waves emitted by a source.
For example, if the source is a spinning neutron star, the frequency of the
gravitational waves it emits can evolve as the frequency of the revolution of
the star.  In general the star will lose its energy and will spin down.  This
evolution of the frequency can be approximated by a Taylor series.  The second
contribution comes from the Taylor expansion of the motion of the Earth around
the Sun.  It is clear that the longer the observation time of the signal the
more terms of the Taylor expansion we need to include in order to accurately
approximate the true signal.

In order to verify the accuracy of the linear model (\ref{Eq:Phs}) we have
calculated the fitting factor FF between the linear model and the {\em accurate} 
model of the signal.  As the accurate model of the phase of the signal we have 
taken the following model:
\be
\label{Eq:Pht}
\Psi_a(t) = \Phi_0 + \omega_d\, t + \sum^{s+1}_{k=1} \omega_k\, t^{k+1}
+ \left[ \omega_0 + \sum^{s+1}_{k=1} (k+1)\omega_k t^k \right]
\frac{{\bf n}_0 \cdot {\bf r}_{\rm SSB}(t)}{c},
\ee
where $\omega_d$ is the signal's frequency $\omega_0$ shifted by a {\em known}
frequency $\omega_s$, i.e.\ $\omega_d=\omega_0-\omega_s$.  The down-conversion
frequency $\omega_s$ (as well as the bandwidth of the signal) can be chosen
arbitrarily and an appropriate time sequence can be extracted from the frequency
domain database as described in Sec.\ \ref{Sec:FDDB}.  The parameters
$\omega_k$ are spin-down parameters and they arise by approximating the 
intrinsic
frequency evolution of the signal by a Taylor expansion.  In the accurate model
(\ref{Eq:Pht}) we include one more spin down than in the linear model
(\ref{Eq:Phs}).  This additional spin down serves to represent the uncertainty
of our model.

For the case when the signal is narrow-band around some frequency $\omega_0$ the
formula (\ref{ff1}) for the fitting factor can be approximated by
\be
\label{ff3}
\text{FF} \approx \max_{\bs{\zeta}}
\tav{ \cos\left[\Psi_a(t;\bs{\zeta}_a)-\Psi_s(t;\bs{\zeta})\right] },
\ee
where 
$\bs{\zeta}_a=(\Phi_0,\omega_0,\omega_1,\ldots,\omega_{s+1},\alpha,\delta)$ are 
the parameters of the accurate model (\ref{Eq:Pht}) and 
$\bs{\zeta}=(p,p_0,p_1,\ldots,p_s,A,B)$ are the parameters of the linear model 
(\ref{Eq:Phs}). The linear phase model (\ref{Eq:Phs}) can shortly be written as
\be
\label{liph}
\Psi_s(t;\bs{\zeta}) = \sum_{i=0}^{s+3} \zeta_i\, l_i(t).
\ee
The phase $\Psi_a$ in the accurate model (\ref{Eq:Pht}) can be expressed as a 
sum similar to the sum from Eq. (\ref{liph}) plus a certain remainder $r$ that 
we assume to be small:
\be
\label{acph}
\Psi_a(t;\bs{\zeta}_a) = \sum^{s+3}_{i=0}
\overline{\zeta}_{ai}(\bs{\zeta}_a)\, l_i(t) + r(t;\bs{\zeta}_a).
\ee
Note that the coefficients $\overline{\zeta}_{ai}$ are some functions of the 
accurate phase parameters $\bs{\zeta}_a$. In numerical calculations described 
below the vector ${\bf r}_{\text{SSB}}$ in the accurate phase model 
(\ref{Eq:Pht}) was computed by our \verb|Top2Bary| code described in Appendix~C, 
and then the coefficients $\overline{\zeta}_{ai}$ and the residual 
$r(t;\bs{\zeta}_a)$ were computed by making a least-squares fit (within the 
observational interval) of the template $\sum_i\overline{\zeta}_{ai}\, l_i(t)$ 
to the accurate phase $\Psi_a(t;\bs{\zeta}_a)$.

Making use of Eqs.\ (\ref{liph}) and (\ref{acph}), from Eq.\ (\ref{ff3}) one 
gets
\bea
\label{ff4}
\text{FF} &\approx& \max_{\bs{\zeta}} \tav{ \cos\left[
\sum^{s+3}_{i=0}\left(\overline{\zeta}_{ai}-\zeta_i\right)l_i(t) + r(t) \right]}
\nonumber\\[2ex]
&\approx& \max_{\bs{\zeta}} \tav{
\left[ 1 - \frac{1}{2} \sum^{s+3}_{i,j=0}
\left(\overline{\zeta}_{ai}-\zeta_i\right)
\left(\overline{\zeta}_{aj}-\zeta_j\right) l_i(t)\, l_j(t) \right] \cos r(t)
- \sum^{s+3}_{i=0} \left(\overline{\zeta}_{ai}-\zeta_i\right) l_i(t) \sin r(t) 
},
\eea
where the last approximation was obtained by Taylor expansion of the cosine 
function and by keeping only the quadratic terms in the (assumed to be small) 
differences $\overline{\zeta}_{ai}-\zeta_i$. It is now easy to find in Eq.\ 
(\ref{ff4}) the maximum over the parameters $\bs{\zeta}$ as the right-hand side 
of Eq.\ (\ref{ff4}) is quadratic in these parameters. The values of the 
parameters that maximize the fitting factor are given by the solution of the 
following set of $s+4$ linear equations:
\be
\label{eqmax}
\sum^{s+3}_{j=0} L_{ij} \left(\overline{\zeta}_{aj}-\zeta_j\right)  = -L_i,
\ee
where
\begin{mathletters}
\bea
L_{ij} &:=& \tav{l_i(t)\, l_j(t) \cos r(t)},
\\[2ex]
L_i &:=& \tav{l_i(t) \sin r(t)}.
\eea
\end{mathletters}

We have used the above prescription to calculate the fitting factor numerically.
Once we have found the parameters of the extremum (\ref{ff4}) by solving Eqs.\
(\ref{eqmax}), we have used these values as input to an optimization routine
(based on Nelder-Mead simplex algorithm) to find an accurate value of the
fitting factor directly from the formula (\ref{ff3}) without the Taylor
expansion. In our calculation we have used the following estimates of the 
maximum values of the spin-down parameters
\be
\label{Mom}
|\omega_k|_{\text{max}} = \frac{\omega_0}{(k+1)\tmin^k},
\ee
where $\tmin$ is the minimum characteristic spin-down age of the neutron star.
These estimates were adopted in the previous papers of this series and they were
taken from the work of Brady {\it et al.}  \cite{BCCS98}.  We have computed the
fitting factor for three phase models with $s=0,1,2$, i.e.\ for a monochromatic
signal, 1-spindown signal, and 2-spindown signal.  We have carried out
computations for two values of the spin-down age ($\tmin=40$~years and
$\tmin=1000$~years), and for different values of the signal's frequency
$f_0=\omega_0/(2\pi)$ and the observation time $T_o$.  For each case we have
made calculation for a grid of positions of the source in the sky.  We have
chosen our observations to start from the beginning of the year 2000 (Julian day
= 2451545.0) and the position of the detector to coincide with the geographical
location of the EXPLORER resonant bar.

Our results are summarized in Tables \ref{Tab:FF1}, \ref{Tab:FF2}, and
\ref{Tab:FF3}.  For each grid of positions of the source in the sky we have
found the minimum value FF$_{\text{min}}$ of the fitting factor (the worst
case).  In the tables we have given the maximum length of the observation time
for which FF$_{\text{min}}$ is greater than $0.9$, $0.9^{1/3}$, and $0.999$. 
The conservative value of the fitting factor equal to $0.9^{1/3} \sim 0.967$ 
comes from the arguments of Apostolatos \cite{A1} by which such 
a fitting factor leads to affordable $10\%$ loss of events. 
The ultraconservative value of the fitting factor equal to $0.999$ 
gives a negligible loss of $0.3\%$ of events. 

{From} the results presented in Tables \ref{Tab:FF1}, \ref{Tab:FF2}, and
\ref{Tab:FF3} it follows that the monochromatic linear model is adequate for a
few hours of the observation time, 1-spindown model for a few days of the
observation time, and 2-spindown model for around 1 week of the observation
time.  For several months of the observation time we do not expect to fit an
adequate linear model however we know that for such long observation times a
coherent all-sky search is computationally prohibitive \cite{BCCS98,JKS98}.
Thus we conclude that in realistic (i.e.\ computationally feasible) coherent
searches of not more than 1 week duration a satisfactory linear model can be
chosen.

\begin{table}
\caption{Adequacy of the monochromatic linear model.}
\label{Tab:FF1}
\begin{center}
\begin{tabular}{ccccccc}
& \multicolumn{6}{c}{Maximal observation time $T_o$ (hours)} \\ \cline{2-7}
Frequency (Hz)
& \multicolumn{3}{c}{$\tmin=40$ years}
& \multicolumn{3}{c}{$\tmin=1000$ years} \\ \cline{2-4}\cline{5-7}
& FF$_{\text{min}}>0.9$
& FF$_{\text{min}}>0.9^{1/3}$
& FF$_{\text{min}}>0.999$
& FF$_{\text{min}}>0.9$
& FF$_{\text{min}}>0.9^{1/3}$  
& FF$_{\text{min}}>0.999$\\ \hline
100 & 4 & 3 & 2 & 8 & 7 & 4\\
200 & 3 & 3 & 1 & 6 & 6 & 3\\
300 & 3 & 2 & 1 & 6 & 5 & 3\\
400 & 2 & 2 & 1 & 5 & 5 & 3\\
500 & 2 & 2 & 1 & 5 & 4 & 3\\
600 & 2 & 2 & 1 & 5 & 4 & 2\\
700 & 2 & 2 & 1 & 5 & 4 & 2\\
800 & 2 & 2 & 1 & 4 & 4 & 2\\
900 & 2 & 2 & 1 & 4 & 4 & 2\\
1000& 2 & 2 & 1 & 4 & 4 & 2\\
\end{tabular}
\end{center}
\end{table}

\begin{table}
\caption{Adequacy of the 1-spindown linear model.}
\label{Tab:FF2}
\begin{center}
\begin{tabular}{ccccccc}
& \multicolumn{6}{c}{Maximal observation time $T_o$ (days)} \\ \cline{2-7}
Frequency (Hz)
& \multicolumn{3}{c}{$\tmin=40$ years}
& \multicolumn{3}{c}{$\tmin=1000$ years} \\ \cline{2-4}\cline{5-7}
& FF$_{\text{min}}>0.9$
& FF$_{\text{min}}>0.9^{1/3}$
& FF$_{\text{min}}>0.999$
& FF$_{\text{min}}>0.9$
& FF$_{\text{min}}>0.9^{1/3}$
& FF$_{\text{min}}>0.999$ \\ \hline
100 & 4 & 3 & 1 & 4 & 3 & 2 \\
200 & 3 & 2 & 1 & 3 & 2 & 1 \\
300 & 2 & 2 & 1 & 3 & 2 & 1 \\
400 & 2 & 2 & 1 & 2 & 2 & 1 \\
500 & 2 & 2 & 1 & 2 & 2 & 1 \\
600 & 2 & 1 & 1 & 2 & 2 & 1 \\
700 & 2 & 1 & 1 & 2 & 1 & 1 \\
800 & 2 & 1 & 1 & 2 & 1 & 1 \\
900 & 1 & 1 & 1 & 2 & 1 & 1 \\
1000& 1 & 1 & 1 & 2 & 1 & 1 \\
\end{tabular}
\end{center}
\end{table}

\begin{table}
\caption{Adequacy of the 2-spindown linear model.}
\label{Tab:FF3}
\begin{center}
\begin{tabular}{ccccccc}
& \multicolumn{6}{c}{Maximal observation time $T_o$ (days)} \\ \cline{2-7}
Frequency (Hz)
& \multicolumn{3}{c}{$\tmin=40$ years}
& \multicolumn{3}{c}{$\tmin=1000$ years} \\ \cline{2-4}\cline{5-7}
& FF$_{\text{min}}>0.9$
& FF$_{\text{min}}>0.9^{1/3}$
& FF$_{\text{min}}>0.999$
& FF$_{\text{min}}>0.9$
& FF$_{\text{min}}>0.9^{1/3}$  
& FF$_{\text{min}}>0.999$ \\ \hline
100 & 14 & 12 & 8 & 14 & 12 & 8 \\
200 & 13 & 11 & 7 & 13 & 11 & 7 \\
300 & 11 & 9  & 6 & 12 & 10 & 6 \\
400 & 10 & 9  & 5 & 10 & 9 & 5 \\
500 & 9  & 8  & 5 & 9  & 8 & 5 \\
600 & 9  & 8  & 5 & 9  & 8 & 5 \\
700 & 9  & 7  & 5 & 9  & 7 & 5 \\
800 & 8  & 7  & 4 & 8  & 7 & 4 \\
900 & 8  & 7  & 4 & 8  & 7 & 4 \\
1000& 8  & 7  & 4 & 8  & 7 & 4 \\
\end{tabular}
\end{center}
\end{table}

\section{Spacing of filters, the search algorithm, and computational 
requirements}
\label{Sec:G}

\subsection{Grid of templates}

In this section we shall present a construction of a grid in the parameter space
on which the statistics $\F$ will be calculated in order to search
for signals.  We assume that as filter (or template) that we use in order to
calculate the statistics we shall use an approximations of the signal by a
suitable linear model studied in Sec.\ \ref{Sec:LM}.  We would like to choose
the grid in such a way that there is no loss of potential gravitational-wave
signals.  In order to determine an appropriate grid of templates we shall use
the correlation function of the statistics $\F$.  We shall choose the
grid in such a way that the correlation function for two filters evaluated for
parameters at two neighboring points of the grid is not less than a certain
specified value.

We thus consider here a constant amplitude signal
\be
\label{gr1}
h\left(t;h_0,\Phi_0,\bs{\xi}\right)
= h_0 \sin\left[ \Phi\left(t;\bs{\xi}\right) + \Phi_0 \right],
\ee
where $\Phi_0$ is the initial phase of the wave form and the vector $\bs{\xi}$ 
collects all other phase parameters. For the signal (\ref{gr1}) the optimal 
statistics $\F$ can be written as
\be
\label{gr2}
\F = \frac{2T_o}{S_h(f_0)}
\left[ \tav{x(t) \cos\Phi(t)}^2 + \tav{x(t) \sin\Phi(t)}^2 \right].
\ee
In Sec.\ IV of Paper III we have shown that when the noise in the detector is a 
zero mean stationary Gaussian process, then the correlation function $C$ 
[defined in Eq.\ (\ref{covdef})] of the random field $\F$, Eq.\ (\ref{gr2}), can 
be approximated by the formula
\be
\label{gr3}
C(\bs{\xi},\bs{\xi}') \approx \tav{\cos[\Phi(t;\bs{\xi})-\Phi(t;\bs{\xi}')]}^2
+ \tav{\sin[\Phi(t;\bs{\xi})-\Phi(t;\bs{\xi}')]}^2.
\ee
When the phase $\Phi$ is a linear function of the parameters $\bs{\xi}$, the 
correlation function $C$ of Eq.\ (\ref{gr3}) depends only on the difference 
$\bs{\tau}:=\bs{\xi}-\bs{\xi}'$ and can be written as
\be
\label{gr4}
C(\bs{\tau}) \approx \tav{\cos[\Phi(t;\bs{\tau})]}^2
+ \tav{\sin[\Phi(t;\bs{\tau})]}^2.
\ee
The fact that for linear phase models the correlation function $C$ depends only 
on the difference $\bs{\xi}-\bs{\xi}'$ between the parameters $\bs{\xi}$ of the 
signal and the parameters $\bs{\xi}'$ of the filter, and not on their absolute 
values, implies that the spacing between the filters will be independent of the 
values of the parameters.  This is one of the advantages of the linear filter 
models.

We further approximate the correlation function $C$ of Eq.\ (\ref{gr4}) by 
Taylor expansion around $\bs{\tau}=0$. Keeping terms at most quadratic in 
$\tau_i$ one gets
\be
\label{nf01}
C\left(\bs{\tau}\right)
\approx 1 - \sum_{i,j} \widetilde{\Gamma}_{ij} \tau_i \tau_j,
\ee
where the matrix $\widetilde{\Gamma}$ has the components 
\be
\label{redgamma}
\widetilde{\Gamma}_{ij}
:= \tav{\frac{\partial\Phi}{\partial\tau_i}\frac{\partial\Phi}{\partial\tau_j}} 
- \tav{\frac{\partial\Phi}{\partial\tau_i}}
\tav{\frac{\partial\Phi}{\partial\tau_j}}.
\ee
The matrix $\widetilde{\Gamma}$ is the reduced Fisher information matrix for
the signal (\ref{gr1}) where the initial phase parameter $\Phi_0$ has been 
reduced (see Appendix B of Paper III for details of this reduction procedure). 

The statistics $\F$ can be expressed as a Fourier transform where the parameter
$p_0$ corresponds to angular frequency and consequently $\F$ can be calculated
using the FFT algorithm.  However the FFT gives the values of the statistics on
a certain grid of frequencies called Fourier frequencies.  These frequencies in
normalized units are separated by factor of $2\pi$.  Thus when the true
frequency falls between the two Fourier frequencies we cannot achieve the
theoretical maximum of the optimal statistics $\F$.  The worst case is when the
frequency falls half way between the Fourier frequencies.  One can easily find
that in such a case the signal-to-noise ratio calculated approximately by means
of the Taylor expansion of the correlation function is equal only to 0.42 of the
optimal signal-to-noise ratio.  This would lead to a drastic loss of signals.
Therefore we need a finer grid.  A way to achieve this and still take advantage
of the speed of FFT is to pad the time series with zeros.\footnote{Padding time
series with zeros of the length of the original time series in real signal
search means that we would need to calculate FFTs of twice the length of the
original data.  This means doubling the computational time and the computer
memory used.  To avoid this pulsar astronomers have invented special
interpolation algorithms that work in the Fourier domain and give twice as fine
Fourier grid from the FFT of the original data.  In the analysis of EXPLORER
data we shall use one such algorithm provided to us by Duncan Lorimer.}  Padding
with zeros of the length of the original time series gives a grid that is twice
as fine as the Fourier grid i.e.\ the difference between the Fourier frequencies
is equal to $\pi$.  Then in the worst case the signal-to-noise ratio is 0.89 of
the optimal one.  As we shall see later from Monte Carlo simulations this choice
of spacing of the frequencies leads to a search algorithm where there is no loss
of events and rms errors of the estimators of parameters are close to 
the theoretical minimum.  
It is also useful to note that the signal-to-noise ratio calculated
form the exact correlation function formula yields the values of the fractions
of the optimal signal-to-nose ratio equal to 0.63 and 0.90 for the difference
between the Fourier frequencies equal to $2\pi$ and $\pi$, respectively.  This
indicates the limits of the validity of the Taylor expansion.

We introduce a convenient normalization of the spin-down parameters $p_k$ of the 
linear model (\ref{Eq:Phs}), namely
\be
\label{gr5}
\overline{p}_k := p_k T_o^{k+1}, \quad k=0,\ldots,s.
\ee
This is equivalent to using a time coordinate $\overline{t}:=t/T_o$ normalized 
by the total observation time $T_o$ (then the normalized time duration of the 
signal is always 1). Using definition (\ref{gr5}) the linear phase model 
(\ref{Eq:Phs}), after dropping the initial phase parameter $p$, can be written 
as:
\be
\label{gr6}
\Phi(t;\bs{\xi}) = \overline{p}_0\, \frac{t}{T_o}
+ \sum_{k=1}^s \overline{p}_k \left(\frac{t}{T_o}\right)^{k+1}
+ A \cos(\Omega_r t) + B \sin(\Omega_r t),
\ee
where $s$ is the number of spin downs included. In the quadratic approximation 
(\ref{nf01}) and for the phase model (\ref{gr6}), equation
\be
\label{gr7}
1 - \sum_{i,j} \widetilde{\Gamma}_{ij} \tau_i \tau_j = C_0 = {\it const},
\ee
describes the surface of the ($s$+3)-dimensional correlation hyperellipsoid.

It is clear that we cannot fill the parameter space completely with
hyperellipsoids.  But such filling can be done by means of some prisms
constructed with the aid of the correlation hyperellipsoid.  We shall illustrate
our construction in the special case of 1-spindown (i.e., $s=1$) linear phase
model (\ref{gr6}).  In this case the components of the matrix
$\widetilde{\Gamma}$ as functions of the observation time are given by [here the
order of the phase parameters is as follows:
$\bs{\tau}=(\Delta\overline{p}_0,\Delta\overline{p}_1,\Delta A,\Delta B)$]
\be
\label{MAT}
{\widetilde \Gamma} = \left(\begin{array}{cccc}
1/12         &1/12           &0               &-1/(2 n\pi) \\[2ex]
1/12         &4/45           &1/(2n^2\pi^2)   &-1/(2 n\pi) \\[2ex]
0            &1/(2n^2\pi^2)  &1/2             &0           \\[2ex]
-1/(2 n\pi)  &-1/(2 n\pi)    &0               &1/2
\end{array}\right),
\ee
where $n$ is the observation time expressed in sidereal days, i.e., 
$n:=(\Omega_r T_o)/(2\pi)$.
 
In the first step we choose spacing in the frequency 
parameter $\overline{p}_0$ to be equal to $\pi$. Then we compute the 
value $C_0$ of the correlation at the surface of the 4-dimensional correlation 
hyperellipsoid (\ref{gr7}) by requesting that the $\overline{p}_0$ axis 
intersects this surface at the points $\Delta\overline{p}_0=\pm\pi/2$. 
Substituting $\bs{\tau}=(\Delta\overline{p}_0,0,0,0)$ into Eq.\ (\ref{gr7}), one 
gets
\be
\label{gr8}
C_0 = 1 - \widetilde{\Gamma}_{\overline{p}_0\overline{p}_0} 
\Delta\overline{p}_0^2 = 1 - \frac{\pi^2}{48} \approx 0.79.
\ee
Making use of Eqs.\ (\ref{MAT}) and (\ref{gr8}), the general equation 
(\ref{gr7}) can be written in the case of 1-spindown signal in the form
[we also substitute $\bs{\tau}=(\overline{p}_0,\overline{p}_1,A,B)$]
\be
\label{gr9}
A^2 + B^2 + \frac{1}{6}\, \overline{p}_0^2
+ \frac{1}{3}\, \overline{p}_0\, \overline{p}_1
+ \frac{8}{45}\, \overline{p}_1^2
+ 2\,\zeta^2 A\, \overline{p}_1
- 2\, \zeta\, B\, (\overline{p}_0+\overline{p}_1)
- \frac{\pi^2}{24} = 0,
\ee
where we have introduced
\be
\label{gr10}
\zeta := \frac{1}{n\pi}.
\ee

We first construct the elementary cell of the grid in the {\em filter space} 
(which is the space spanned by the parameters $\overline{p}_1$, $A$, and $B$).
We consider the $\overline{p}_0=0$ cross section of the 4-dimensional 
hyperellipsoid (\ref{gr9}). This cross section defines the 3-dimensional 
ellipsoid in the $(\overline{p}_1,A,B)$ space:
\be
\label{gr11}
(A+\zeta^2\overline{p}_1)^2 + (B-\zeta\,\overline{p}_1)^2
= \frac{\pi^2}{24} - \left(\frac{8}{45}-\zeta^2-\zeta^4\right) \overline{p}_1^2
\ee
(let us note that $\frac{8}{45}-\zeta^2-\zeta^4>0$ for $n\ge1$).
Next we take the cross sections of the ellipsoid (\ref{gr11}) with the planes 
$\overline{p}_1=\pm\delta$ for some positive constant $\delta$. These cross 
sections define two adjacent circles. We inscribe two regular hexagons into 
these circles. The hexagons form the bases of the inclined prism inscribed in 
the ellipsoid (\ref{gr11}). The 3-dimensional volume of this prism reads
\be
\label{gr12}
V(\delta) = 3\sqrt{3} \left[\frac{\pi^2}{24} - 
\left(\frac{8}{45}-\zeta^2-\zeta^4\right) \delta^2 \right] \delta.
\ee
Then we maximize the volume $V(\delta)$ with respect to $\delta$. The function 
$V(\delta)$ attains its maximal value for
\be
\label{gr13}
\delta_{\text{max}}
= \frac{\pi}{6\sqrt{2}\sqrt{\dst \frac{8}{45}-\zeta^2-\zeta^4}}.
\ee
The maximal value $V(\delta_{\text{max}})$ of the volume (\ref{gr12}) is the 
volume of the elementary cell in the filter space. It reads
\be
\label{gr14}
V_{\text{gr}}
= \frac{\pi^3}{24\sqrt{6}\sqrt{\dst \frac{8}{45}-\zeta^2-\zeta^4}}.
\ee
Equation (\ref{gr14}) gives the following values of the volume $V_{\text{gr}}$ 
of the elementary cell in the filter space:
$V_{\text{gr}}=2.05$, 1.35, 1.29, 1.27 for $n=1$, 2, 3, 4, respectively.

The elementary cell in the 4-dimensional space
$(\overline{p}_0,\overline{p}_1,A,B)$ we construct as follows.  The
3-dimensional prism of maximal volume, described above, which lies in the
$\overline{p}_0=0$ subspace, we parallelly translate in four dimensions, in two
opposite directions, into the subspaces $\overline{p}_0=-\pi/2$ and
$\overline{p}_0=+\pi/2$.  As the direction of translation we choose one of the
principal directions\footnote{The reasonable result one obtains only when
translating along this principal direction of the hyperellipsoid (\ref{gr9}),
which almost coincides with the $\overline{p}_0$ axis.}  of the hyperellipsoid
(\ref{gr9}).  Such constructed elementary cell is the 4-dimensional prism which
bases are two adjacent 3-dimensional hexagonal prisms lying in the
$\overline{p}_0=-\pi/2$ and $\overline{p}_0=+\pi/2$ subspaces.

It is clear from the construction that our elementary cell will stick out the
4-dimensional hyperellipsoid (\ref{gr9}).  For the case of $n=2$ we have
calculated the correlation function
$C(\bs{\tau})=1-\sum_{i,j}\widetilde{\Gamma}_{ij}\tau_i\tau_j$ at all vertices
of the elementary cell and we have found that the smallest value of
$C(\bs{\tau})$ is equal to 0.77.  Thus the grid constructed above ensures that
the correlation between the filter and the signal is not less than 0.77.  The
Monte Carlo simulations presented below show that using such a grid we do not
lose any signals and that rms errors of the parameter estimators are very close
to the minimum ones allowed by the Cram\`er-Rao bound.

If as a base of the elementary cell in the filter space we choose a square
instead of a regular hexagon the volume of the cell (independently of the value
of $n$) decreases by a factor of $3\sqrt{3}/4 \sim 1.3$.

\subsection{Monte Carlo simulations}

In order to test the effectiveness of the chosen grid we have made Monte Carlo
simulations of the detection of our signal in the noise and estimation of
its parameters. In the simulations we have taken the signal $s$ to be
1-spindown linear model with a constant amplitude $h_0$:
\be
\label{Eq:1s}
s(t) = h_0 \exp \left\{ i \left[
p + p_0\, t +  p_1\, t^2 + A \cos\left(\frac{2\pi n t}{T_o}\right)
+ B \sin\left(\frac{2\pi n t}{T_o}\right) \right] \right\},
\ee
where $T_o$ is the observation time and $n$ is the integer equal to the number 
of sidereal days of observation in real search. In the case of signal 
(\ref{Eq:1s}) the maximum likelihood detection involves finding the global 
maximum  of the functional $\F_s$ with respect to the parameters $p_0$, $p_1$, 
$A$, $B$. The functional $\F_s$ is given by
\be
\label{mc1}
\F_s = \frac{2|\widetilde{X}|^2}{S_h T_o},
\ee
where
\be
\label{mc2}
\widetilde{X} = \int_0^{T_o} x(t)
\exp \left\{ -i \left[ p_1\, t^2 + A \cos\left(\frac{2\pi n t}{T_o}\right)
+ B \sin\left(\frac{2\pi n t}{T_o}\right)
\right] \right\} \exp(-ip_0t)\, dt.
\ee
In Eq.\ (\ref{mc2}) the data $x(t)=s(t)+n(t)$, where $n(t)$ is the stationary
noise with spectral density $S_h$.  In our simulations we have generated white
and Gaussian noise $n(t)$.  Thus our statistics $\F_s$ consists of taking the
modulus of the Fourier transform of the data demodulated for a grid of
parameters $p_1$, $A$, and $B$.  To find the maximum of $\F_s$ we have used a
hierarchical algorithm consisting of two steps:  a \emph{coarse} search and a
\emph{fine} search.  The coarse search involves calculation of $\F_s$ on the
grid in the parameter space constructed in Sec.\ VII A and finding the maximum
value of $\F_s$ on that grid.  The values of the parameters of the filter that
give the maximum are coarse estimates of the parameters of the signal.  The fine
search involves finding the maximum of $\F_s$ using a maximization routine with
the starting values of the parameters equal to the coarse estimates of the
parameters.  As a maximization routine we have used the Nelder-Mead simplex
algorithm.  This algorithm involves only the calculation of the function to be
maximized at certain points but not its derivatives.  We plan to use the above
hierarchical procedure in analysis of real data from the EXPLORER detector.

The procedure described above differs from another two-step hierarchical
algorithm proposed by Dhurandhar and Mohanty \cite{DM96,Moh98}.  The first step
of the two procedures is the same but in the second step Dhurandhar and Mohanty
propose to use a fine grid in the parameter space around the maximum obtained
from the coarse search whereas we propose to use a maximization routine.

We have performed Monte Carlo simulations by generating a signal $s(t)$ given by
Eq.\ (\ref{Eq:1s}) and adding white Gaussian noise $n(t)$.  By adjusting the
amplitude $h_0$ we have generated a signal with a chosen optimal signal-to-noise
ratio $d$.  In our simulations we have chosen the observation time to be $n=2$.
We have done 1000 runs for several values of $d$.  We have compared the standard
deviations of the parameter estimators obtained from the simulations with the
theoretical ones calculated from the Cram\`er-Rao bound.  We have also compared
the probability of detection of the signal obtained from the simulations with
the theoretical one calculated from Eq.\ (\ref{PD}).  In our simulations we have
used the hexagonal grid constructed in Sec.\ \ref{Sec:G} A.

\begin{figure}
\begin{center}
\scalebox{0.89}{\includegraphics{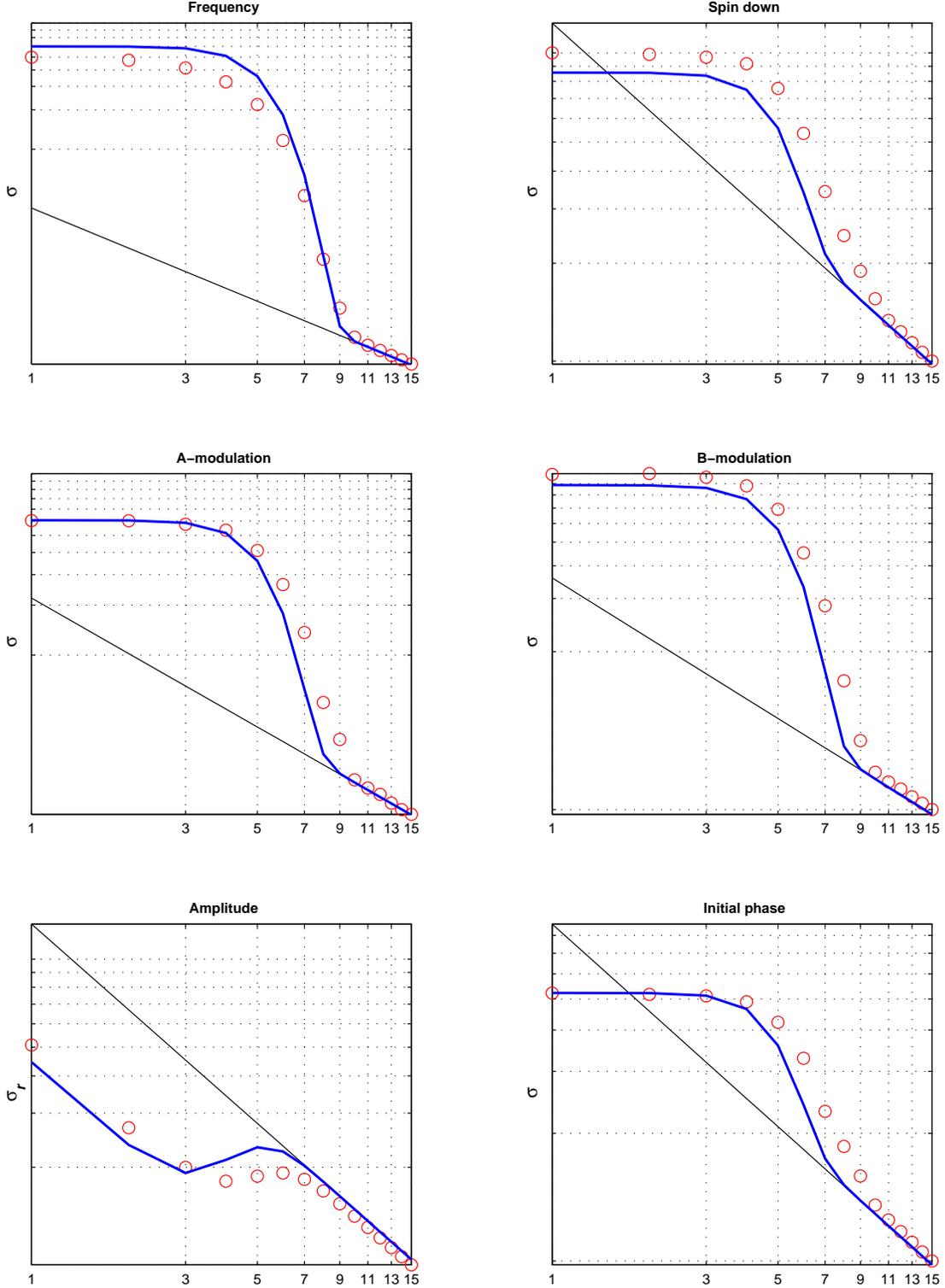}}
\end{center}
\caption{Monte Carlo simulations of the rms errors of the parameter estimators 
of the signal given by Eq.\ (\ref{Eq:1s}) for $n=2$. The $x$ axes are labeled 
by the optimal signal-to-noise ratio $d$. The $y$ axes are labeled by the 
standard deviation $\sigma$ except for the case of the amplitude parameter
where the relative rms error $\sigma_r:=\sigma/h_0$ is given. The results of 
simulations of 1000 runs for each value of $d$ are marked by the circles.  The 
thin solid lines are calculated from covariance matrix and they constitute the 
Cram\`er-Rao bound, i.e.\ the smallest rms error for an unbiased estimator.  The 
thick solid lines are calculated from a model that takes into account the false 
alarms.}
\label{crb2ab}
\end{figure}

The results of our computations are depicted in Fig.\ \ref{crb2ab}.  We have
observed that above a certain signal-to-noise ratio (that we shall call the {\em
threshold signal-to-noise ratio}) the results of the Monte Carlo simulations
agree very well with the calculations of the rms errors from the covariance
matrix.  However below the threshold signal-to-noise ratio they differ by a
large factor.  This threshold effect is well-known in signal processing
\cite{VT69} and has also been observed in numerical simulations for the case of
a coalescing binary chirp signal \cite {KKT94,BSD96}.  As was explained in Sec.\
VII of Paper III this effect arises because sometimes the global maximum of the
functional $\F_s$ occurs as a result of noise and not the signal.  This happens
the more often the lower the signal-to-noise ratio.  Following the theory of
this effect developed in Paper III we have calculated the approximate rms errors
of the estimators of the parameters.  They are shown as thick lines in Fig.\
\ref{crb2ab}.

The comparison of probability of detection obtained from the simulations with
the theoretical formula (\ref{PD}) shows that for lower signal-to-ratios we have
more detections than expected.  This is because the theoretical formula  
assumes that when the signal is detected its parameters are located in the cell
corresponding to the true parameters of the signal.  However in practice for
lower values of $d$ as a result of the noise the signal may drift to neighboring
cells.  Above the threshold signal-to-noise ratio we find that simulated 
probability of detection almost exactly agrees with the theoretical one 
and that we are not losing any signals.

\subsection{Estimation of parameters of the signal}

Once the optimal statistics, calculated with the linear filter of Sec.\
\ref{Sec:LM}, crosses a chosen threshold we register the estimates of the
parameters $p_0$, $p_1$, $A$, and $B$.  We would like to estimate physically
interesting parameters:  frequency, spin down, declination and right ascension
of the source.  We propose the following algorithm to achieve this.  {From}
Eqs.\ (\ref{deal}) and least-squares fit of the linear model to the accurate
model of the signal we calculate the approximate values of parameters
$\omega_0$, $\omega_1$, $\delta$, $\alpha$.  Then using the accurate model of
the signal as a filter we search for accurate estimates of the parameters on a
small grid around the approximate estimates.

We have made Monte Carlo simulations of the above procedure by making 100 runs
for one position of the source in the sky and for five signal-to-noise ratios:
8, 12, 16, 20, 24.  First of all we have found that probability of detection of
the signal agrees very well with the theoretical one and thus we are not losing
any signals.  Concerning accuracy of the parameter estimation we have obtained
that rms errors for declination and right ascension were very close to their
Cram\`er-Rao bounds, however the rms errors for frequency and spin down were
worse.  The mean biases in the estimators of the parameters $\omega_0$,
$\omega_1$, $\delta$, and $\alpha$ were less than 0.1\%, 0.5\%, 0.005\%, and
0.01\%, respectively.

\subsection{Amplitude modulation}

Approximation of the continuous gravitational-wave signal by a simple linear
model and consequently approximation of the optimal statistics, Eq.\ (\ref{OS}),
by a homogeneous random field were studied under the assumption that the
amplitude of the signal was constant.  However as the modulation of the
amplitude of the signal is very slow (it changes on a time scale of one sidereal
day) in comparison to the rapid modulation of the phase we expect that our
linear model of the phase is a satisfactory approximation of the phase of the
optimal filters given by Eqs.\ (\ref{Fab}).  We also expect that the calculation
of the number of elementary cells [by means of Eq.\ (\ref{NT}); except for a
factor of 2, see below] and construction of the grid in the parameter space
given in Sec.\ VII A above are valid for the amplitude modulated signal.

To calculate the optimal statistics, Eq.\ (\ref{OS}), we first need to correct
the time series for amplitude modulation i.e.\ multiply the time series by the
functions $a(t)$ and $b(t)$.  Our grid is on the space parameterized by $p_0$,
$p_1$, $A$, and $B$ whereas the functions $a(t)$ and $b(t)$ depend on the
declination $\delta$ and the right ascension $\alpha$.  From Eqs.\ (\ref{AB}) we
have the following expressions for $\delta$ and $\alpha$ in terms of $A$ and
$B$:
\begin{mathletters}
\label{deal}
\bea
\delta &=& \pm \arccos \left(
\frac{\sqrt{A^2+B^2}}{\dst \frac{\omega_0 r}{c}} \right),
\\[2ex]
\alpha &=& \phi_r + \arctan\left(\frac{A}{B}\right).
\eea
\end{mathletters}
Since for each set of parameters $A$ and $B$ there are two sets of parameters 
$\delta$ and $\alpha$  the number of cells that we obtain from Eq.\  (\ref{NT})
needs to be multiplied by a factor of 2 and this number is to be used to
compute from Eq.\ (\ref{FP}) the threshold corresponding to a given false alarm 
probability.

We see that the equation for declination $\delta$ of the source depends on the
angular frequency $\omega_0$ of the signal that we do not know before we
detected the signal.  The uncertainty in $\delta$ affects the constant
amplitudes in front of the time dependent modulation factors.  However for the
case of the EXPLORER detector data (which we plan to analyze employing the
methods developed above) we have that frequency $f_0$ is around 1~kHz within the
band $B\sim1$~Hz (see Sec.\ \ref{Sec:EXPs} below).  Thus by choosing in the
equation for declination $\delta$ the frequency equal to the middle frequency of
the band to be analyzed instead of the unknown frequency $\omega_0$, we shall
lose only a fraction $\sim{B/(2f_0)}\sim0.0005$ of the signal-to-noise ratio.

\subsection{Computational requirements}

To estimate the computational requirements to do the search we
calculate the number of FFTs that one needs to compute. This number depends on 
the total volume $V_F$ of the filter space, which is given by
\be
\label{vtaf}
V_F = \int\!\!\int_{B_2(0,r_o)} dA\,dB
\int_{-\omega_{\text{max}}T_o^2/{(2\tmin)}}^{\omega_{\text{max}}T_o^2/{(2\tmin)}
} d\,\overline{p}_1,
\ee
where $\omega_{\text{max}}$ is the maximum angular frequency of the signal and
$B_2(0,r_o)$ is a 2-dimensional disc in the $(A,B)$ plane of radius $r_o$,
\be
r_o = \frac{\omega_{\text{max}}r}{c}.
\ee
Thus we get
\be
V_F = \frac{\pi}{c^2}
\frac{\omega_{\text{max}}^3 r^2 T_o^2 }{\tmin}.
\ee

The number of grid points on which the optimal statistics $\F$, Eq.\ (\ref{OS}),
should be calculated is obtained by dividing the volume $V_F$ of the filter
space by the volume $V_{\text{gr}}$ of the elementary cell of the chosen grid in 
the filter space.  In the previous subsection we have found that for each set
of parameters $A$ and $B$ there are two sets of parameters $\delta$ and $\alpha$
[see Eqs.\ (\ref{deal})].  Moreover the optimal statistics involves calculations
of two FFTs --- $F_a$ and $F_b$ [see Eqs.\ (\ref{OS}) and (\ref{Fab})].  Thus
the total number $N_{\text{FFT}}$ of FFTs is 4 times the number of grid points
and it is given by
\be
\label{NFFT}
N_{\text{FFT}} = 4\frac{V_F}{V_{\text{gr}}}.
\ee

Assuming that the data processing rate should be comparable to the data 
acquisition rate and that the data processing consists only of computing
the FFTs the number ${\mathcal P}$ of floating point operations per second
(flops) can be estimated by
\be
\label{Nflops}
{\mathcal P} = 6\Delta\nu N_{\text{FFT}} [\log_2(2\Delta\nu T_o) + 1/2],
\ee
where $\Delta\nu=(\omega_{\text{max}}-\omega_s)/(2\pi)$ is the bandwidth of the 
search.

\section{A narrowband all-sky search of EXPLORER data}
\label{Sec:EXPs}

The data analysis tools for searching continuous gravitational-wave signals that
we have developed in the previous sections of the present work can be applied
both to the resonant bar and the laser interferometric detectors.  We plan to
apply these tools to the data from the resonant bar detector
EXPLORER\footnote{The EXPLORER detector is operated by the ROG collaboration
located in Italian Istituto Nazionale di Fisica Nucleare (INFN); see
\verb|http://www.roma1.infn.it/rog/explorer/explorer.html|.} \cite{exp}.  The
detector has already collected many years of data with high duty cycle (e.g.\ in
1991 the duty cycle was 75\%).  Our primary objective is to carry out an all-sky
search.  It is a unique property of the gravitational-wave detectors that with a
single time series one can search for signals coming from all sky directions.
In the case of other instruments like optical and radio telescopes to cover the
whole or even part of the sky requires a large amount of expensive telescope
time.  The directed search of the galactic center has already been carried out
and limits for the amplitude of the gravitational waves has been established
\cite{ROGc00}.

The EXPLORER detector is most sensitive over certain two narrow bandwidths
(called minus and plus modes) of about 1~Hz wide at two frequencies around 
1~kHz.
To make the search computationally feasible we propose an all-sky search of data
of a few days long in the narrow band were the detector has the best
sensitivity.  By narrowing the bandwidth of the search we can shorten the length
of the data to be analyzed as we need to sample the data at only twice the
bandwidth and thus we reduce the computational time.  To reduce the parameter
space to search we restrict ourselves to only one spin-down parameter.  We would
also like to use the linear filter model as the search template.  Then as our
frequency is around 1~kHz from Table \ref{Tab:FF2} we read that we can consider
up to 2 days of coherent observation time in order that the fitting factor is
greater then 0.9.  For the sake of the FFT algorithm it is best to keep the
length of the data to be a power of 2.  Consequently we choose the number of
data points to analyze to be $N=2^{18}$.  Then for $T_o=2$ days of observation
time the bandwidth $\Delta\nu$ of the data will be
$\Delta\nu=N/(2T_o)\sim0.76$~Hz.  We also choose to analyze the data for the
plus mode which has frequency around 922~Hz.  With these parameters we can
calculate the number of filters that we need to compute.  {From} Eq.\
(\ref{NFFT}) and for the hexagonal grid over the sky we get that
$N_{\text{FFT}}\sim3.7\times10^8$.  Assuming that the data processing rate
should be comparable to the data acquisition rate the computing power required
[calculated by means of Eq.\ (\ref{Nflops})] is around 7.7 Gflops.  If we allow
a month for off-line processing of data with the above parameters we only need
around 250 Mflops of computer power.

Using Eq.\ (\ref{NT}) we can calculate the number of cells $N_c$ in the
parameter space.  Then from the number of cells, using Eq.\ (\ref{sFP}) we can
calculate the threshold that we need to set.  For example choosing the false
alarm probability to be 1\% we find by inverting Eq.\ (\ref{sFP}) and using Eq.
(\ref{EF}) that for the parameters that we have chosen above the threshold
signal-to-noise ratio is 8.3.  However we know that in the coarse search we are
using a suboptimal filter and we are losing signal-to-noise ratio.  For the
parameters of our search we find that in the worst case the fitting factor is
0.94.  Moreover due to our discrete grid in the parameter space the
signal-to-noise ratio can decrease in the worst case by an additional factor of
$\sqrt{0.78}\sim 0.88$.  Thus in the worst case the signal-to-noise ratio of our
coarse search can be a factor of $0.83$ of the optimal one.  Hence in order not
to lose any signals we lower the signal-to-noise threshold by a factor of 0.83.
By lowering the threshold we must make sure that the number of false alarms does
not increase excessively.  {From} Eq.\ (\ref{sNF}) we find that the expected
number of false alarms with the lower threshold is $\sim 310$.  This is
certainly a manageable number that will insignificantly increase the
computational time of the total search.  We can consider lowering the threshold
even further in order to have more candidate events that can be later verified
using longer stretches of data.  For example by lowering the threshold by a
factor of 0.8 the expected number of false alarms is around 1600 which will
still be manageable to verify.

The minimum detectable amplitude $h_0$, i.e.\ the amplitude for which the 
signal-to-noise ratio is equal to 1, is given by
\be
h_0 = \sqrt{\frac{S_h}{T_o}},
\ee 
where $S_h$ is the {\em one-sided} spectral density of noise.  For 2 days of
observation time $T_o$ and spectral density $S_h$ at the plus mode equal to
$2\times10^{-42}/\text{Hz}$, the minimum detectable amplitude is
$3.4\times10^{-24}$.  In order that we have a detection with 99\% confidence by
the calculation above we need a signal-to-noise ratio of 8.3 and thus the
amplitude of the signal must be around $2.8\times10^{-23}$.  Consequently by
estimate given in Eq.\ (\ref{hon}) a continuous gravitational-wave signal from a
neutron star located at a distance of 1~kpc, spinning with period of 2~ms, and
with ellipticity $\epsilon\sim10^{-5}$ will be detectable.

\section*{Acknowledgments}

One of us (A.K) would like to thank ROG group in Istituto Nazionale di Fisica 
Nucleare (INFN), Frascati, where part of this work was done, for hospitality 
and Consiglio Nazionale delle Ricerche (CNR) for support within the exchange
agreement with Polish Academy of Sciences.  This work was supported in part by
the KBN Grant No.\ 2 P03B 094 17 (K.M.B., P.J., and A.K.).  We would also like
to thank Interdisciplinary Center for Mathematical and Computational Modeling of
Warsaw University for computing time.

\appendix

\section{Derivation of the bar beam-pattern functions}

We consider here the response of a bar detector to a weak plane gravitational 
wave in the long wavelength approximation. Moreover, we assume that the 
frequency spectrum of the gravitational wave which hits the bar entirely lies 
within the sensitivity band of the detector. Under these assumptions the 
dimensionless response function $h$ of the bar detector can be computed from
the formula (cf.\ Sec.\ 9.5.2 in Ref.\ \cite{T87})
\be
\label{apa1}
h(t) = {\bf n}\cdot\left[\widetilde{H}(t){\bf n}\right],
\ee
where ${\bf n}$ denotes the unit vector parallel to the symmetry axis of the 
bar, $\widetilde{H}$ is the three-dimensional matrix of the spatial metric 
perturbation produced by the wave in the proper reference frame of the detector, 
and a dot stands for the standard scalar product in the three-dimensional 
Cartesian space. In the detector's reference frame we introduce Cartesian 
detector coordinates $(x_{\text{d}},y_{\text{d}},z_{\text{d}})$ with the 
$z_{\text{d}}$ axis along the Earth's radius pointing toward zenith, and the 
$x_{\text{d}}$ axis along the bar's axis of symmetry. In these coordinates 
the vector ${\bf n}$ from Eq.\ (\ref{apa1}) has components
\be
\label{apa2}
{\bf n}=\left(1,0,0\right).
\ee
In the wave Cartesian coordinate system 
$(x_{\text{w}},y_{\text{w}},z_{\text{w}})$ (in which the wave travels in the 
$+z_{\text{w}}$ direction), the three-dimensional matrix $H$ of the wave induced 
spatial metric perturbation has components
\be
\label{apa3}
H(t) = \left(\begin{array}{ccc}
h_+(t) & h_{\times}(t) & 0
\\[2ex]
h_{\times}(t) & -h_+(t) & 0
\\[2ex]
0 & 0 & 0
\end{array}\right),
\ee
where the functions $h_+$ and $h_\times$ describe two independent wave's
polarizations. The matrices $\widetilde{H}$ and $H$ are related through equation
\be
\label{apa4}
\widetilde{H}(t) = M(t)\,H(t)\,M(t)^{\text{T}},
\ee
where $M$ is the three-dimensional orthogonal matrix of transformation from the
wave coordinates to the detector coordinates, T denotes matrix transposition. 
Collecting Eqs.\ (\ref{apa1})--(\ref{apa4}) together one can see that the 
response function $h$ is a linear combination of the functions $h_+$ and 
$h_\times$:
\be
\label{apa5}
h(t)=F_+(t)h_+(t)+F_{\times}(t)h_{\times}(t),
\ee
where $F_+$ and $F_\times$ are beam-pattern functions.

Because of the diurnal motion of the Earth the beam-patterns $F_+$ and
$F_\times$ are periodic functions of time with a period equal to one sidereal
day.  We want to express $F_+$ and $F_\times$ as functions of the right
ascension $\alpha$ and the declination $\delta$ of the gravitational-wave source
and the polarization angle $\psi$ (the angles $\alpha$, $\delta$, and $\psi$
determine the orientation of the wave reference frame with respect to the
celestial reference frame defined below).  We represent the matrix $M$ of Eq.\
(\ref{apa4}) as
\be
\label{apa6}
M = M_3\,M_2\,M_1^{\text{T}},
\ee
where $M_1$ is the matrix of transformation from wave to celestial frame
coordinates, $M_2$ is the matrix of transformation from celestial coordinates to
cardinal coordinates and $M_3$ is the matrix of transformation from cardinal
coordinates to detector coordinates.  In celestial coordinates the $z$ axis
coincides with the Earth's rotation axis and points toward the North pole, the
$x$ and $y$ axes lie in the Earth's equatorial plane, and the $x$ axis points
toward the vernal point.  In cardinal coordinates the $(x,y)$ plane is tangent
to the surface of the Earth at detector's location with $x$ axis in the
North-South direction and $y$ axis in the West-East direction, the $z$ cardinal
axis is along the Earth's radius pointing toward zenith.  Under the above
conventions the matrices $M_1$, $M_2$, and $M_3$ are as follows (matrices $M_1$
and $M_2$ are taken from Sec.~II~A of Ref.\ \cite{JKS98})
\begin{mathletters}
\label{apa7}
\bea
\label{m1}
M_1 &=& \left(\begin{array}{ccc}
\sin\alpha\cos\psi-\cos\alpha\sin\delta\sin\psi&
-\cos\alpha\cos\psi-\sin\alpha\sin\delta\sin\psi&
\cos\delta\sin\psi
\\[2ex]
-\sin\alpha\sin\psi-\cos\alpha\sin\delta\cos\psi&
\cos\alpha\sin\psi-\sin\alpha\sin\delta\cos\psi&
\cos\delta\cos\psi
\\[2ex]
-\cos\alpha\cos\delta&
-\sin\alpha\cos\delta&
-\sin\delta
\end{array}\right),
\\[2ex]
\label{m2}
M_2 &=& \left(\begin{array}{ccc}
\sin\phi\cos(\phi_r+\Omega_r t)&
\sin\phi\sin(\phi_r+\Omega_r t)&
-\cos\phi
\\[2ex]
-\sin(\phi_r+\Omega_r t)&
\cos(\phi_r+\Omega_r t)&0
\\[2ex]
\cos\phi\cos(\phi_r+\Omega_r t)&
\cos\phi\sin(\phi_r+\Omega_r t)&
\sin\phi
\end{array}\right),
\\[2ex]
\label{m3}
M_3 &=& \left(\begin{array}{ccc}
-\sin\gamma & \cos\gamma & 0
\\[2ex]
-\cos\gamma & -\sin\gamma & 0
\\[2ex]
0 & 0 & 1
\end{array}\right).
\eea
\end{mathletters}
In Eq.\ (\ref{m2}) $\phi$ is the geodetic latitude of the detector's site,
$\Omega_r$ is the rotational angular velocity of the Earth, and the phase
$\phi_r$ defines the position of the Earth in its diurnal motion at $t=0$ (the
sum $\phi_r+\Omega_r t$ essentially coincides with the local sidereal time of 
the
detector's site; see Sec.\ VII~B).  In Eq.\ (\ref{m3}) $\gamma$ determines the
orientation of the bar's axis of symmetry with respect to local geographical
directions, it is measured counter-clockwise from East to the bar's axis of
symmetry.

To find the explicit formula for $F_+$ and $F_\times$ we have to combine Eqs.\ 
(\ref{apa1})--(\ref{apa7}). After some algebraic manipulations we arrive at
the expressions (\ref{bpb}) and (\ref{abb}).
Equivalent explicit formulas for the functions $a$ and $b$ from Eqs.\ 
(\ref{abb}) can be found in Ref.\ \cite{M98} where different
angles describing the position of the gravitational-wave source in the sky and
the orientation of the detector on the Earth are used.\footnote{Our functions 
$a$ and $b$ from Eqs.\ (\ref{abb}) are identical with the 
functions ${\cal S}_+$ and ${\cal S}_\times$ from Eqs.\ (2.90) and (2.91) of 
Ref.\ \cite{M98}, provided the following identification of the variables $\eta$, 
$\alpha$, $\theta$, $\Psi$ used in \cite{M98} with our variables $\Omega_r$, 
$\phi_r$, $\alpha$, $\delta$, $\phi$, $\gamma$ is made:
$\eta\to-(\alpha-\phi_r-\Omega_r t)$,
$\alpha\to\pi/2-\delta$,
$\theta\to\pi/2-\phi$,
$\Psi\to\gamma-\pi/2$.}

\section{The time averages $\tav{\lowercase{a}^2}$ and $\tav{\lowercase{b}^2}$}

\subsection{Laser interferometeric detector}

The time averages $\tav{a^2}$ and $\tav{b^2}$ entering Eqs.\ (\ref{amle0}), for 
the observation time $T_o$ chosen as an integer number of sidereal days, for the 
laser interferometeric detector take the form (here $n$ is a positive integer):
\begin{mathletters}
\bea
\tav{a^2}\bigg\vert_{T_o=n\,2\pi/\Omega_r}
&=& \frac{1}{16} \sin^22\gamma \left[
9\cos^4\phi\cos^4\delta
+ \frac{1}{2}\sin^22\phi\sin^22\delta
+ \frac{1}{32}\left(3-\cos2\phi\right)^2\left(3-\cos2\delta\right)^2 \right] 
\nonumber\\[2ex]&&
+ \frac{1}{32} \cos^22\gamma \left[ 4\cos^2\phi\sin^22\delta
+ \sin^2\phi\left(3-\cos2\delta\right)^2 \right],
\\[2ex]
\tav{b^2}\bigg\vert_{T_o=n\,2\pi/\Omega_r}
&=& \frac{1}{32} \sin^22\gamma \left[
\left(3-\cos2\phi\right)^2\sin^2\delta + 4\sin^22\phi \cos^2\delta \right] 
\nonumber\\[2ex]&&
+ \frac{1}{4} \cos^22\gamma \left(1 + \cos2\phi\cos2\delta\right).
\eea
\end{mathletters}

We see that $\tav{a^2}$ and $\tav{b^2}$ depend only on one unknown parameter of
the signal --- the declination $\delta$ of the gravitational-wave source.  They
also depend on the latitude $\phi$ of the detector's location and the
orientation $\gamma$ of the detector's arms with respect to local geographical
directions.

\subsection{Resonant bar detector}

For the resonant bar detector the time averages $\tav{a^2}$ and $\tav{b^2}$, for 
the observation time $T_o$ being an integer number of sidereal days, have the 
form:
\begin{mathletters}
\bea
\tav{a^2}\bigg\vert_{T_o=n\,2\pi/\Omega_r} 
&=& \frac{1}{4} \left(1-3\sin^2\gamma\cos^2\phi\right)^2 \cos^4\delta
+ \frac{1}{2} \sin^2\gamma\cos^2\phi \left(1-\sin^2\gamma\cos^2\phi\right) 
\sin^22\delta
\nonumber\\[2ex]&&
+ \frac{1}{8} \left[\sin^22\gamma\sin^2\phi + 
\left(\cos^2\gamma-\sin^2\gamma\sin^2\phi\right)^2 \right]
\left(1 + \sin^2\delta \right)^2,
\\[2ex]
\tav{b^2}\bigg\vert_{T_o=n\,2\pi/\Omega_r}
&=& \frac{1}{2} \left( \sin^22\gamma\cos^2\phi + \sin^4\gamma\sin^22\phi 
\right) \cos^2\delta
\nonumber\\[2ex]&&
+ \frac{1}{2} \left( \cos^4\gamma + \frac{1}{2}\sin^22\gamma\sin^2\phi
+ \sin^4\gamma\sin^4\phi \right) \sin^2\delta.
\eea
\end{mathletters}

As in the case of the laser interferometeric detector, the time averages
$\tav{a^2}$ and $\tav{b^2}$ depend on the declination $\delta$ of the
gravitational-wave source, the latitude $\phi$ of the detector's location, and
the orientation $\gamma$ of the detector's axis of symmetry with respect to
local geographical directions.

\section{The Top2Bary procedure}
\label{EPH}

The algorithms described in Sec.\ V were implemented in an easily callable 
subroutine named \verb|Top2Bary| consisting of about 900 lines of FORTRAN code.  
The ephemeris files (\verb|DE405'90.'10|, \verb|tai-utc.dat| and yearly 
\verb|eopc04.yy|) required by the program must be kept in the directory 
\verb|\Top2Bary\EphData| on the current disk.  The procedure header is shown 
below.

\begin{verbatim}
 subroutine Top2Bary(CJD_E,clat,clong,height,pve,pvo)

c Procedure to calculate position (km) and velocity (km/s) of an observatory
c located at given geographical position (conventional coordinates in degrees
c and height above the Earth ellipsoid in m) at given Julian day.
c It requires some ephemeris files, placed in the 'path' directory.

c        Argument description:
c CJD_E - Julian Day number representing the Coordinated Universal Time
c         If CJD_E is negative it is assumed that it is -1*(Ephemeris JD).
c clat  - observatory conventional geographical latitude (deg).
c clong - observatory conventional geographical longitude (deg).
c height- observatory height above the Earth ellipsoid (m).
c pvo   - 6 element array of barycentric vectors (3 for position in km,
c         and 3 for velocity, km/s) of the observatory relative to Earth baryc.
c pve   - 6 element array of barycentric vectors (3 for position in km,
c         and 3 for velocity, km/s) of the Earth barycenter relative to SSB.

c More important subroutines called:
c  polmot(CJD,Px,Py,UT1_UTC,dpsi,deps) - reads polar motion and UT1 data
c  topobs(DJ1,glat,glong,height,pvo,amst) - returns observer's coord. in 'pvo'
c  earthPV(BJD,pve) - returns Earth barycentric position (J2000 frame) in 'pve'
\end{verbatim}

\end{document}